\documentclass[smallextended]{svjour3}
\renewcommand{\makeheadbox}{}
\AtBeginDocument{
  }

\usepackage{graphicx}
\usepackage{booktabs}
\usepackage{algorithm} 
\usepackage{xurl}
\usepackage{hyperref}
\usepackage{orcidlink}
\usepackage{amsmath}
\usepackage{float}
\usepackage{enumitem}
\usepackage[most]{tcolorbox}
\usepackage{multirow}
\usepackage{rotating}
\usepackage{wrapfig}
\usepackage{svg}
\usepackage{amsmath}
\newcommand{\xquote}[2]{\textcolor{gray}{``\textit{#1}"}$_{#2}$}

\usepackage{xcolor}
\smartqed

\definecolor{fillyellow}{HTML}{FED894}
\definecolor{outlineyellow}{HTML}{FCC416}
\definecolor{likertleft}{HTML}{304A47}
\definecolor{likertright}{HTML}{3D0B37}
\definecolor{findings}{HTML}{be9a2d}
\usepackage{fontspec}
\newfontface\carlitofont{Carlito-Regular.ttf}

\usepackage{tikz}
\newcommand*\circled[1]{
    \tikz[baseline=(char.base)]{
        \node[
            shape=circle,
            draw=outlineyellow,   % Outline color
            fill=fillyellow,      % Fill color
            text=black,           % Text color
            inner sep=0pt,
            font=\small\carlitofont     % Use Calibri font within the circle
        ] (char) {#1};
    }
}
\definecolor{greyblue}{HTML}{8497B0}
\newcommand{\dv}[1]{\textcolor{greyblue}{#1}}

\newtcolorbox{mybox}[1]{colback=outlineyellow!5!white,colframe=fillyellow!75!black,fonttitle=\bfseries,title=#1, breakable}

\newcommand{\sysmsgicon}{
  \raisebox{-0.2\height}{\includegraphics[height=1em]{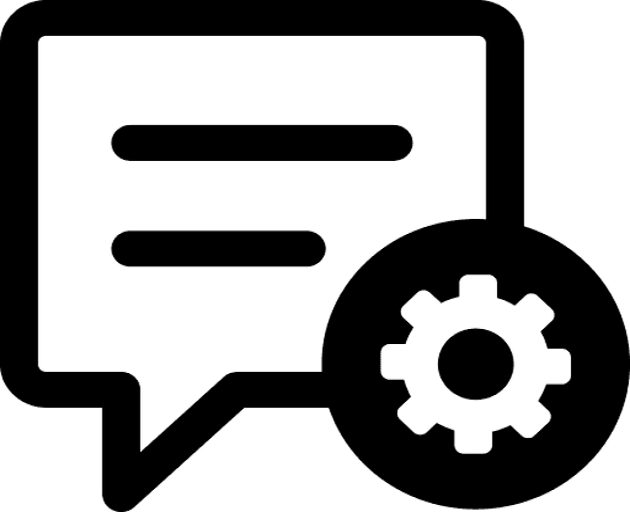}}
}

\newcommand{\usericon}{
  \raisebox{-0.2\height}{\includegraphics[height=1em]{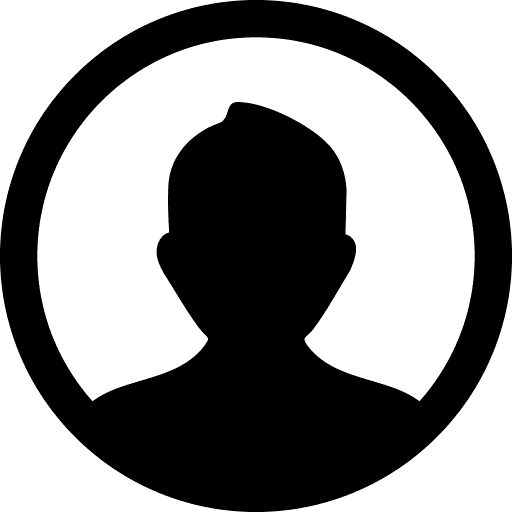}}
}

\newcommand{\boticon}{
  \raisebox{-0.2\height}{\includegraphics[height=1em]{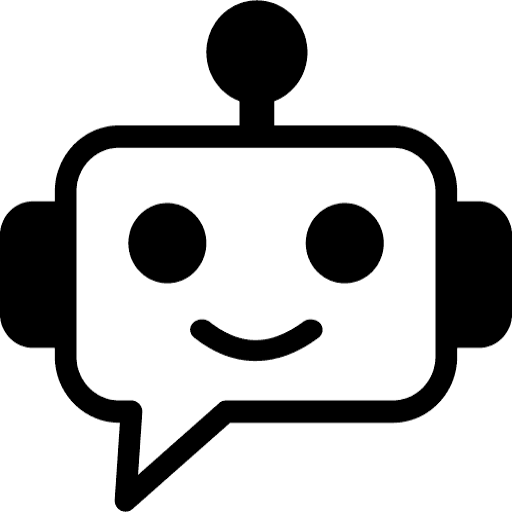}}
}

\newlist{questions}{enumerate}{2}
\setlist[questions,1]{label=\textbf{RQ$_\text{\arabic*.}$},ref=RQ$_\text{\arabic*}$, leftmargin=*}
\setlist[questions,2]{label=\textbf{(\alph*)},ref=\thequestionsi(\alph*), leftmargin=*}

\newcommand{\ques}[1]{\texorpdfstring{RQ$_{#1}$}{RQ #1}}

\newcommand{\context}[0]{\emph{context}}
\newcommand{\control}[0]{\emph{control}}
\newcommand{\proactive}[0]{\emph{proactive}}

\let\oldcite\cite
\renewcommand{\cite}{\,\oldcite}

\begin{document}

\newpage

\title{On the Prospects of Dynamic LLM Conversations in Software Development}

\titlerunning{On the Prospects of Dynamic LLM Conversations in Software Development}

\author{Annemarie Wittig\orcidlink{0000-0002-7051-2242} \and Alina Mailach\orcidlink{0000-0001-6204-2095} \and Janet Siegmund\orcidlink{0000-0002-5815-266X} \and Norbert Siegmund\orcidlink{0000-0001-7741-7777}}

\institute{A. Wittig \at
        Leipzig, Leipzig University, Leipzig, 04109, Germany \\
        \email{annemarie.wittig@cs.uni-leipzig.de}           
    \and
        A. Mailach \at
        Leipzig, Leipzig University, Leipzig, 04109, Germany \\
    \and
        J. Siegmund \at
        Chemnitz University of Technology, Chemnitz, 09107, Germany   \\
    \and
        N. Siegmund \at
        ScaDS.AI Dresden/Leipzig, Leipzig University, Leipzig, 04109, Germany
}
\date{}
\authorrunning{Annemarie Wittig et al.}

\maketitle

\begin{abstract} 
Large language models (LLMs) have become an essential tool for assisting developers, yet we still lack knowledge on ways to effectively support their interactions during development activities. That is, the quality of interactions with a chat-based LLM still strongly depends on how developers phrase prompts and which information they include.

Our goal is to evaluate whether interventions into these interactions with LLMs have an effect on software developers---be it harmful or beneficial. To this end, we conducted a four-month longitudinal study with third-semester computer science students working on a full-stack Web development project using chat-based LLMs under three conditions: (1) a \emph{context}-aware group received intent-based conversation augmentation, (2) a \emph{proactive} group received follow-up suggestions and tailored advice, and (3) a \emph{control} group without intervention.
Our augmentations are minimal: (i) to reduce confounding factors and (ii) to isolate treatment effects.

Analyzing interaction logs and user surveys revealed no major differences in interaction patterns, indicating no detectable harmful effects in the measured outcomes when intervening in interactions. Moreover, we observed trends of increased satisfaction with the \emph{proactive} treatment. The results indicate that even with minimal interventions, dynamic guidance mechanisms for developer-LLM interactions show observable effects, such that more severe augmentations may have the potential to substantially improve developer satisfaction.
\end{abstract}

\section{Introduction}
The growing popularity of large language models (LLMs) has made them essential to assist software engineers in their daily work. These models have proven useful for many software engineering tasks, including code generation\cite{codegen,codegendebugging,debugging,codecommentgen}, requirements elicitation\cite{requirementelicitation}, debugging\cite{codegendebugging,debugging}, and documentation\cite{littleimprovement,codecommentgen,impactsllms}. 

\setlength{\columnsep}{7pt}
\begin{wrapfigure}[24]{r}{0.37\textwidth}
    \vspace{-1.5em}
    \centering
    \includegraphics[width=0.33\textwidth]{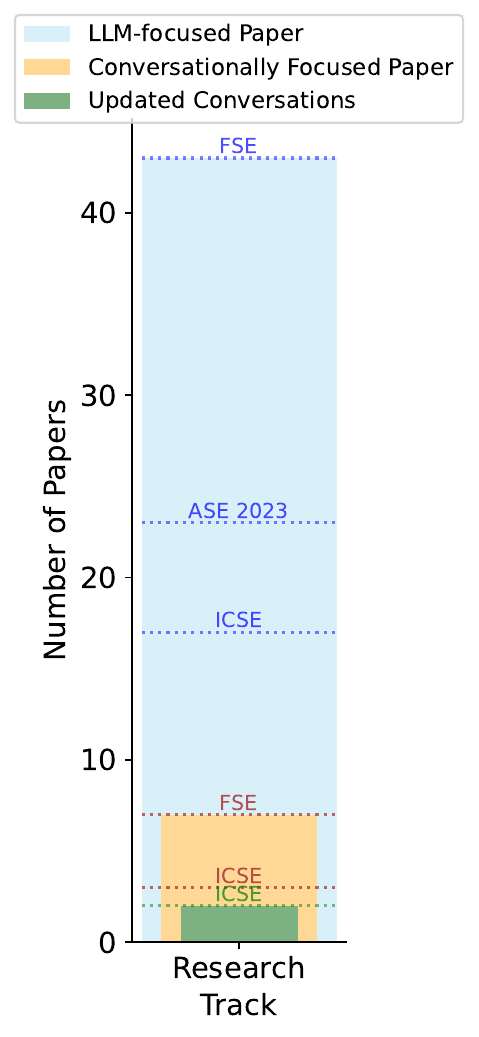}
    \vspace{-0.5em}
    \caption{Number of research papers on LLMs published in FSE'24, ICSE'24, and ASE'23}
    \label{fig:papers}
\end{wrapfigure}

A number of studies have been conducted that focus on how developers rely on LLM-based tools, investigating usage and interactions\cite{surveyonusage}, and the impact on productivity: 
Findings regarding the usefulness are mixed, with some studies finding no\cite{surveyonproductivity}, little\cite{littleimprovement}, or considerable improvement\cite{copilotcasestudy}.
Despite such numerous research activities, only few papers focused on how to guide users through the interaction with an LLM to increase their usefulness.
\autoref{fig:papers} shows that, among 43 LLM-centered papers presented at the research tracks of recent ICSE, ASE, and FSE conferences, only seven analyze the conversational structure and developer interactions with LLMs. 
Just two of these papers attempt to improve these interactions. 
This underscores a critical gap: While LLMs are widely applied, the central question of how users interact with them and how to better support this interaction remains underexplored. 
Closing this gap is essential to unlock the full potential of LLM-developer interactions.

The lack of research prioritizing the guidance of developers in their conversations with LLMs is surprising, as it has been shown in different research fields that supporting chat conversations of individuals can yield more positive results for both sides of the conversation, for example, by causing a better understanding of conversational contexts\cite{procrastinationstudy,heartrate,groupbot,empathy}. Moreover, studies in software engineering found that developers often do not provide sufficient context to an LLM, reducing its ability to effectively solve a development task\cite{newton,littleimprovement,mailach,llmplugin}. Despite the advances in prompting techniques, retrieval augmented generation, and agentic development, there is limited insight into how conversations may benefit from additional guidance.
Thus, we ask whether and how guided developer–LLM conversations benefit software development, and how they affect developer behavior and perceptions as well as the resulting code.

In this study, we investigate the effects of two approaches intended to support developer-LLM interactions:  The first approach, referred to as \emph{context treatment}, enhances the system message provided to the LLM with context based on a user-selected intention chosen before each conversation. 
The second treatment, referred to as \textit{proactive treatment}, provides additional guidance by recommending improved and follow-up prompts, as well as advice tailored to the user's conversation.
We show such a workflow in \autoref{fig:reply}, highlighting the different mechanisms of each treatment.

Our goal is to understand how these direct interventions into interactions with an LLM can affect software developers and the software development process. To this end, we evaluate both treatments against a control group in a large study with computer science students in their 3rd semester working on a \textit{full-stack web development project} for \textit{four consecutive months}.
This study design, focusing on an entire project rather than a small programming task, and lasting for four months instead of a few hours, is a distinguishing feature of this work. 
It enables us to draw long-term effects in a more practical setting. Despite relying on students, our setting has high ecological validity due to the size, complexity, variety, and length of the task.
We provided developers with a user interface to interact with the adjusted LLM and logged all interactions.
Additionally, we conducted a monthly survey tracking the user's experience throughout the experiment.

Overall, our treatments are designed to follow a conservative approach with a focus on internal validity, so we explore whether even minimal interventions can already affect the development process. Although we do not expect large effect sizes with such a setup, we thereby minimizing the risk of introducing too many confounding factors (e.g., the way interventions are provided or whether they disturb the LLM interaction) and can better pinpoint effects to certain treatments. Based on this controlled design, we provide the foundation for research on extensive and targeted interventions to unlock further productivity gains with LLMs.
The main goal of the study is to identify whether and to what degree even small interventions in the communication of an LLM affect development.

In summary, we make the following contributions:
\begin{itemize}
    \item A description of two novel approaches to guide developers during software development with LLMs.
    \item An analysis of longitudinal usage data from three different LLM assistants (\emph{control}, \context{}, \proactive{}) over a four-month, realistic software development project.
    \item A comprehensive replication package containing an application with both guidance approaches and all material and data from the empirical study.\footnote{\url{https://anonymous.4open.science/r/On-the-Prospects-of-Dynamic-LLM-Conversations-in-Software-Development-D22C}}
\end{itemize}

\section{Related Work}

\paragraph{Pitfalls of LLM Usage.}
Several studies have observed that LLMs misunderstand what users want and need to successfully achieve their goal. For instance, LLMs misunderstand requirements\cite{littleimprovement,surveyonusage}, task context\cite{surveyonusage}, and problems, goals, and rationales resulting in faulty conversation\cite{surveyonproductivity,surveyonusage}.
Particularly in more complex software engineering tasks, LLMs appear to be less beneficial compared to simple coding tasks\cite{rockscoding}.  
These studies provide valuable insights in the context of developer-LLM interactions, but none of these take a close look at how we can guide developers to overcome these challenges---the key motivation for our work. Thus, we focus 
on automatically intercepting in developers' interactions to provide situation-specific support in complex software engineering projects.

\paragraph{Improving LLM Interactions for Development.}

Several studies focus on finetuning LLMs by incorporating specific information sources to enhance LLM capabilities.
Kong et al.\cite{selfprompttuning} use pre-collected data to fine-tune an LLM, such that it automatically refines the latest prompt of a conversation using role prompting~\cite{kong2024better}, a technique that assigns a model a role via prompts to guide its responses in line with that role's expertise and perspective. While  promising, updating only the latest prompt in a conversation ignores the conversational flow that chatbot interactions usually have. 
Ouyang et al.\cite{supervisefinetune} fine-tuned LLMs with human feedback to better align them with user intent, which results in improved truthfulness and less toxic replies. 
Similarly, Wang\cite{reinforcementlearning} applied reinforcement learning from human feedback to adapt interactions. 
While these studies explore similar directions to ours, they directly update models, which reduces flexibility and ties them to specific model instances. Moreover, a model update is static by definition and cannot account for new conversational intents and contexts.
By contrast, our approach is dynamically intercepting and updating conversations while they are ongoing, allowing for larger flexibility with less time and compute.

Another approach to improve user-LLM interaction is to intervene directly in conversations to enhance the interaction by including specific context information.
Nam et al.\cite{llmplugin} enhance open-ended prompts with user-highlighted code, resulting in more task completions. They conclude that adding even more contextual information holds potential to improve interactions.
Arteaga Garcia et al.\cite{newton} validates this conclusion by demonstrating that, with more context, such as environment variables, previously executed code, and previous chat messages, the LLM yields better results.   
Similarly, Li et al.\cite{contextide} add IDE native contexts, hereby focusing on code completion instead of conversations.
While these studies follow partially related ideas, they do not concentrate on the prompt flow, but include different context information. A distinguishing feature of our study is that it is not being limited to coding tasks, specific IDEs, and laboratory environments, but aims at analyzing effects on intervening chat interactions in a long-term, practical development scenario. That is, it is orthogonal to the aforementioned studies.

\paragraph{Guidance of Chat Interactions in Different Fields.}
The need for supporting chat interactions is not limited to SE contexts, but has been recognized in multiple different fields.
Researchers enhanced chatbots for more empathetic and efficient responses for embodied conversational agents in general conversations to better react and understand participants intents\cite{empathy}.
Another study on LLMs for procrastination management found that LLMs require nuanced contexts to properly react to user's needs which are usually not provided, thus recommends the inclusion of automatically generated dynamic contexts\cite{procrastinationstudy}.
Overall, even if the approaches differ, research fields beyond SE recognized that conversations can benefit from additional support and guidance to improve the overall experience and effectiveness.

\section{Methodology}
In this section, we present our experimental design by introducing and motivating the research questions, then operationalizing the independent and dependent variables, and finally detailing the qualitative analysis, participants, and materials.

\subsection{Research Questions}
\label{sec:rq}

The overarching goal of this research is to understand the effect of guidance mechanisms on developers and the code they produce. 
Specifically, we are interested in how different levels of guidance affect developers’ interactions with an LLM, their perception of the LLM, and the code they produce. To this end, we define three research questions:

\begin{questions}
	\item How does the level of guidance influence developer behavior when interacting with an LLM?
	\item How do developers perceive LLMs with guided conversations?
	\item How do different guidance and support mechanisms influence code produced by developers?
\end{questions}

We address these questions through three dependent variables: {user behavior}, {perception}, and {generated code usage}, each capturing interaction, experience, and code artifacts, respectively.

\paragraph{Independent Variable.}

\begin{figure}
	\centering
    \includegraphics[width=1\textwidth]{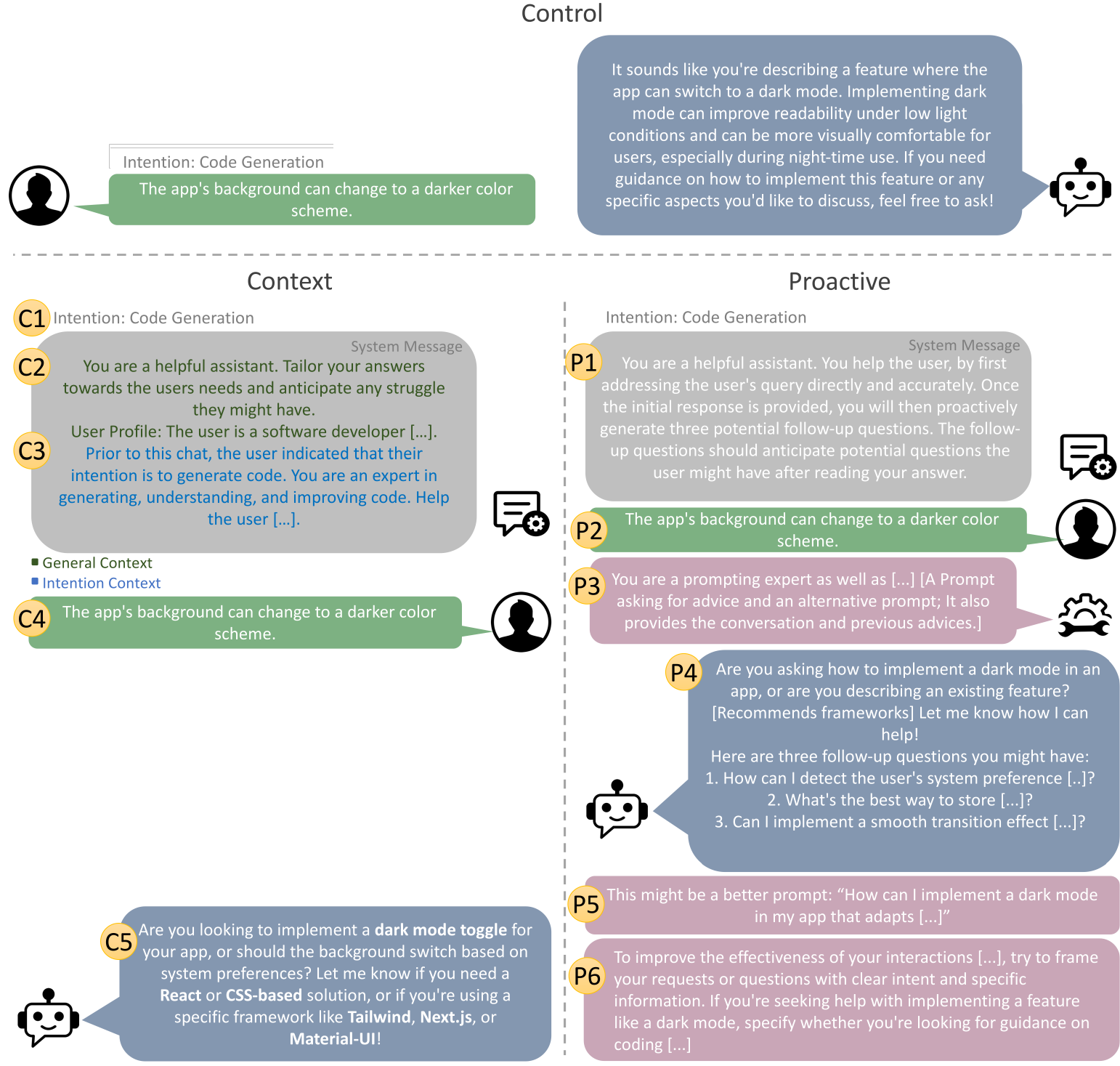}
	\caption{The system message (\protect\sysmsgicon\ gray bubble), developer input (\protect\usericon\ green bubble), bot response (\protect\boticon\ blue bubble), and how they are updated in each treatment, using the example input prompt ``The app's background can change to a darker color scheme.'', representing a minimal feature description}
    %\Description{Workflow showing how the input `Generate me a markdown heading' is treated each treatment group, and how the replies might look for those.} 
	\label{fig:reply}
\end{figure}

The level of guidance is the independent variable and has three alternatives: \dv{contextual guidance}, \dv{proactive guidance}, and \dv{no guidance} (our control condition). \autoref{fig:reply} illustrates how these levels play out in a conversation with the same user prompt.

We operationalize \dv{contextual guidance} (lower left part in \autoref{fig:reply}), such that we enhance user-conversations with the context of the project task given by us (i.e., developing a web application) \circled{C2} and the developer's intention for a conversation \circled{C3}, which they provide at the start of every conversation by selecting from a list of pre-defined intentions or providing their own \circled{C1}. This way, we ensure sufficient context for the LLM, so that it can generate more tailored responses, for example, to either provide code, a description, or a user story. We chose this treatment because prior research has shown that providing sufficient context to an LLM is essential for good developer–LLM interactions: 
Challenges often arise when developers do not share the necessary context\cite{Davila2024AnIC,khojah2024codegenerationobservationalstudy,mailach}, leading to  frustration\cite{khojah2024codegenerationobservationalstudy}, out-of-context interactions\cite{Davila2024AnIC}, and incorrect or unsatisfying replies\cite{mailach}. 
Furthermore, since we found in a previous study that participants have difficulty providing sufficient context\cite{mailach}, we decided to automatically provide the context to the LLM without participant interaction. This allows us to focus on the effect of additional context, independent of participants' ability to provide it. Related approaches have also been successfully explored by other researchers\cite{newton,contextide,llmplugin}.

Next, we operationalize \dv{proactive guidance} (lower right part in \autoref{fig:reply}) by instructing the LLM to act as a helpful assistant and predict follow-up questions that the user may state based on the generated answer \circled{P1}. Additionally, we give the user feedback on the quality of their prompts and hints for more effective interactions with the LLM \circled{P3} (based on a hidden LLM). Inspired by automatic prompt refinement techniques\cite{selfrefinement}, we use the reply to suggest a better suited prompt to the user in an informative block \circled{P5}, alongside the generated hints and advice \circled{P6}. This operationalization is also inspired by our previous study, in which participants also faced difficulties in formulating prompts\cite{mailach}. Furthermore, our experience shows that participants find it difficult to navigate programming projects and determine which direction to take next. Thus, we included this treatment to provides support in these areas.

For the control condition, \dv{no guidance}, we use a LLM without any intervention (upper part in Figure~\ref{fig:reply}).
We refer to this model as the control LLM to ensure clear separation from the treatment conditions.
Both treatments are minimally invasive compared to the control LLM, for two reasons: First, we focus on internal validity to understand possible effects in detail. Thus, if the treatments differ only in few aspects, we can more reliably attribute cause and effect relationship between treatments and outcome. Second, we reduce ethical issues to a minimum. If we design a treatment that is highly invasive and worsens the performance of students, the respective students would be disadvantaged disproportionally.

We provide a short summary of each of the levels of guidance in \autoref{tab:independent}.

\begin{table}[ht]
    \centering
    \caption{Overview of our levels of guidance and their operationalization}
    \begin{tabular}{l p{0.32\textwidth} p{0.32\textwidth}}
    \toprule
        Level of Guidance & Operationalization & Rationale \\ \midrule
        No Guidance & Base LLM, no intervention & Control Group \\
        Contextual Guidance & Enhanced user conversations with task and intention context & Mitigation of challenges caused by LLMs lacking context \\
        Proactive Guidance & Additional follow-up questions, feedback on prompt quality and further hints, alternative prompts & Mitigation of issues with prompt formulation and finding directions in programming projects \\ \bottomrule
    \end{tabular}
    \label{tab:independent}
\end{table}

\paragraph{Dependent Variables.} 

To answer our RQs, we analyze 3 dependent variables, one per RQ: \dv{user behavior}, \dv{perception}, and \dv{generated code usage}. We provide a summary of each in \autoref{tab:dependent} .

The outcome \dv{user behavior} describes the interaction of developers with the LLM  and represents the most direct observable impact of the treatments. 
We operationalized it three-ways.
First, we checked the \emph{conversation length} (number of prompts per conversation, their evolvement over time, and number of prompt tokens, i.e. tokens from user-created messages, per conversation) to measure how extensively participants interacted with the LLM and whether this changed over time or under treatment.
Second, we evaluated the \emph{specified intentions} for starting a conversation, provided by the participants, to examine whether treatments steered developers toward switching their intention (e.g. explanations versus code generation).
Third, we analyzed the \emph{intention alignment} of participants' provided intention with conversation content. This allows us to explore underlying reasons of potential treatment impacts.
To assess whether participants' specified intentions align with the prompt contents within conversations, we selected 45 conversations per treatment to inspect manually.
The resulting sampling size of 135 conversations was chosen to ensure sufficient coverage across treatments while keeping manual inspection feasible.
We used stratified random sampling, such that we order the conversations by time, assign them to buckets of 100 conversations, and randomly selected five conversations per treatment from each bucket. 
Then, two of the authors assigned each of these prompts the intention that best fit the content of the prompt.
We compared whether our assignments align with the user-specified intention and logged each mismatch.

\dv{Perception} refers to \emph{perceived helpfulness} of the LLM, its perceived impact on developers' \emph{frustration}, and their \emph{productivity}. We assessed all three aspects with monthly questionnaires. This provides insights into how participants perceive guided interactions, and determines whether these guidance mechanisms could serve as a foundation for future research on improving developer-LLM interactions.

Last, we observe the \dv{generated code usage} for the projects. Specifically, we analyze the \emph{amount of generated code} pushed into a repository, the \emph{modification rate} of committed code, and the \emph{duration} for how long this code typically remains in the project.
We collected this data to analyze how much the LLM supports development efforts, how long-lasting this support is, and how our treatment changes these effects.
To this end, we randomly selected 50 conversations from each treatment for manual review. 
We restricted the selected conversations to intentions that tend to naturally produce code: \textit{code generation}, \textit{language question}, \textit{debugging}, and \textit{test generation}. A conversation must also contain more than two prompts, and at least one response with generated code that not just repeats code of the prompt and does not solely contain shell commands or settings. For each conversation, we reviewed the commit history of the corresponding repository from the day the code was generated to the final release tag. When the generated code was indeed committed, we analyzed the commit history and counted line modifications to evaluate how the code evolved.

\begin{table}[ht]
    \centering
    \caption{Dependent Variables, the RQ for which they are collected, and their operationalization.}
    \begin{tabular}{p{2cm}p{3cm}p{5cm}}
    \toprule
        \textbf{Variable} & \textbf{Operationalization} & \textbf{Rationale} \\
        \midrule
        \ques{1}: \newline {User Behavior }
            & Conversation length & Changes in interaction intensity under treatment and time\\
            & Specified intentions & Treatment influence on the type of interaction developers initiate \\
            & Intention alignment & Exploration of reasons for treatment impacts \\
        \midrule
        \mbox{\ques{2}: Developer}\newline {Perception }
            & Helpfulness, frustration, productivity & Participants' perception of guided interactions\\
        \midrule
        \mbox{\ques{3}: Generated} Code Usage 
            & Amount pushed & Adoption rate of AI-generated code in development\\
            & Modification rate & Degree of adaptation required for AI-generated code\\
            & Duration & Persistence of AI-generated code in the codebase\\
        \bottomrule
    \end{tabular}
    \label{tab:dependent}
\end{table}

\subsection{Study Conduct}

We conducted the study in the context of a mandatory 3rd semester software engineering course offered at two universities. This course included practical work to develop a full-stack web development project that lasted four months. 
The project was either preceded by a lecture on software engineering (one university) or accompanied by one (second university), covering the software engineering lifecycle, development practices, and prompt engineering basics.

With the start of the project, participants received access to the chatbot that was connected to GPT-4\cite{openai2024gpt4}. 
Using the most recent model at the time ensured participants' interest in the LLM while reducing the risk of abandonment due to perceived tool limitations.
They first gave their informed consent, confirming that they want to participate in the study, and agreed to the pseudonymized use of their data. 
They could revoke these agreements and stop participating in the study at any time without any repercussions on their course fulfillment.
Once consent was given, participants could use the chatbot throughout the four-month project while completing their regular course activities.
Throughout this time, we collected participant information through questionnaires, including short monthly questionnaires to collect longitudinal data on their opinions and satisfaction. Participants could discontinue the study at any time without any negative impact. In that case, they would have received access the chatbot using an account without tracking and without questionnaire prompts. No participant withdrew consent or requested removal of their data. There was no ethics proposal, which is not required at either university; we followed standard research guidelines on these kinds of studies.

\begin{figure*}
    \centering
    \includegraphics[width=1\textwidth]{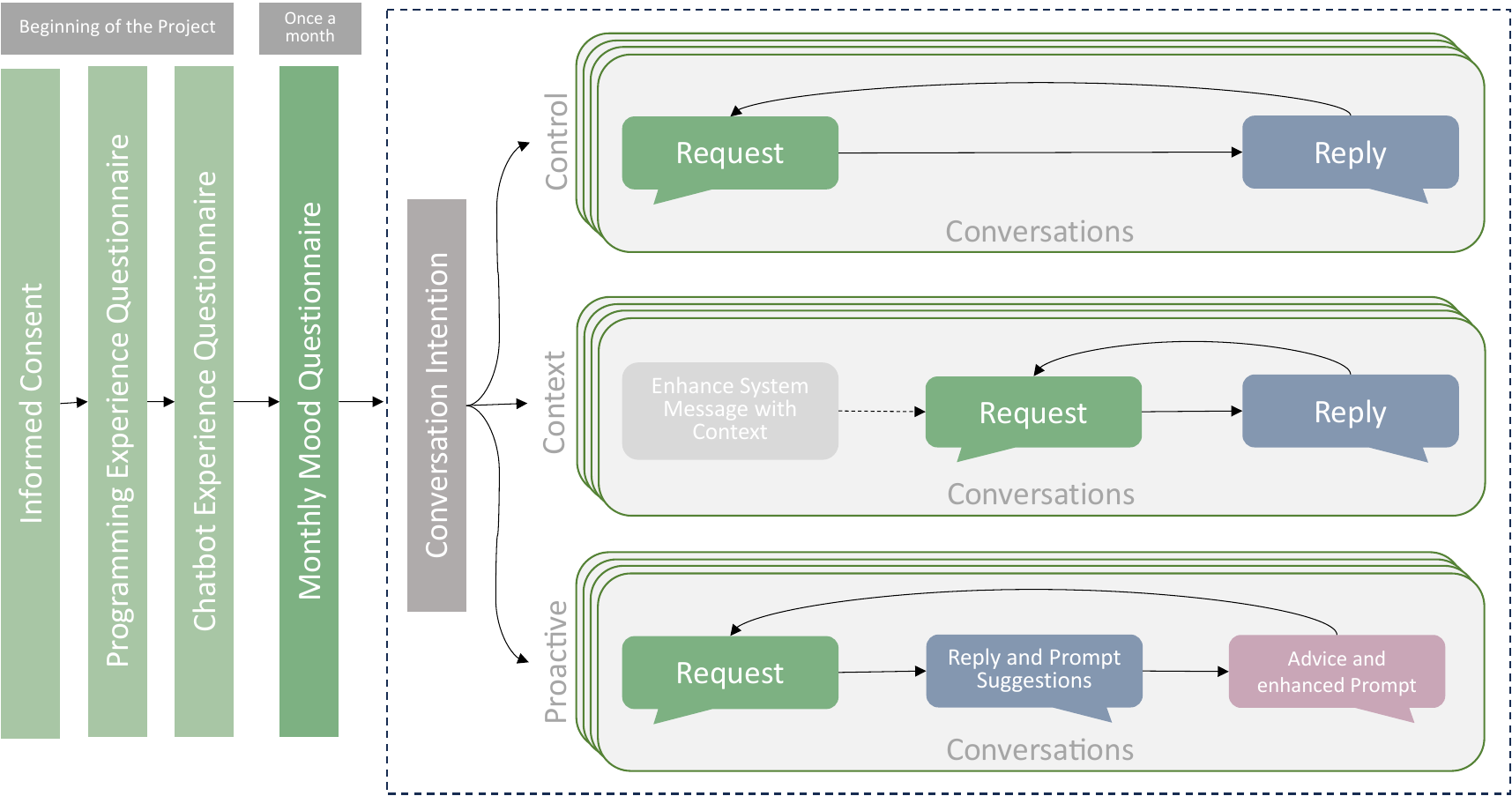}
    \caption{Process of the study after participants received access depending on the treatment}
    \label{fig:overview}
\end{figure*}

There was no restriction in using the chatbot. 
At the beginning of each conversation, we asked our participants to specify the intention they have and whether they use the LLM for the project or other use cases. 
The study process for participants is visualized in \autoref{fig:overview}.
We analyze only project-related conversations.
We restrict the chatbot history to read-only. 
This ensures the validity of our experiment by maintaining the relevance of intentions to the current conversation.

\subsection{Material}
\label{sec:material}

\paragraph{Intentions.}
When initializing a conversation with the LLM, participants were asked to specify their intention for this conversation. 
The intention choice cannot be changed within a conversation, as it is intended that participants start new conversations when their intentions change.
This ensures that conversations are clearly assigned to intentions, prioritizing internal validity. 
This process is the same for all treatments.

For starting a conversation, selecting one of the following intentions was necessary:
\textit{to generate code} (referred to as \textit{code generation}, \textit{code gen}), \textit{to debug code} (\textit{debugging}, \textit{debug}), \textit{to test code} (\textit{test generation}, \textit{test gen}), \textit{to have code explained} (\textit{code comprehension}, \textit{code comp}), \textit{to identify requirements} (\textit{requirements elicitation}, \textit{req elic}), \textit{to ask questions about lecture content} (\textit{lecture question}, \textit{lect ques}), and lastly, \textit{to ask questions about the programming language/framework} (\textit{language question}, \textit{lang ques}).
The specification of a custom intention was possible if none of these were suitable.
The intention pool has been derived from our previous study by Mailach et al.\cite{mailach}.

\paragraph{Questionnaires.}
\label{material:questionnaire}
During the course of the study, we applied several questionnaires to answer our research questions.
At the beginning, we asked participants about their programming knowledge, following the guidelines by Siegmund et al.\cite{siegmund}. 
 Details are in the replication package.

Additionally, we assessed information about experience with prompt engineering (PE), including whether they have attended a dedicated PE lecture of the course.
Questions included how often participants have used LLMs, whether previous experience was positive or negative, and whether they would like to use them in the future. In a free text field, participants could also share any thoughts that they might feel relevant for the study.

The monthly surveys assessed participants' perception of the LLM to answer our second research question.
Completing these questionnaires was mandatory to continue interacting with the LLM. 
These surveys asked, on a 5-point-Likert scale, whether they find the chatbot helpful, frustrating, and beneficial for productivity. 
The monthly questionnaire also included an optional open-text field for any thoughts participants liked to share with us.

We chose a one-month timeframe to collect regular insights and track changes throughout the experiment, while accommodating the typical bursts and pauses of activity in university projects. 
This duration also minimized participant inconvenience, reducing the risk of participants stopping using the chatbot due to excessive interruptions.

\subsection{Participants}
\label{sec:analysis}
\label{preprocessing}
To work on the programming project, students were randomly divided into teams of 4 to 8 students.
The teams were then randomly assigned to the levels of the independent variable, ensuring similar experiences, even when multiple participants interacted with one LLM, for example, in the context of pair programming.
In total, 93 students started the experiment. 
For answering \ques{1} and \ques{3}, we cleaned the data to remove outliers in three steps:
\begin{enumerate}
    \item Removing all conversations / prompts with a non-project intention (removed 7 participants from the dataset).
    \item Removing participants who sent fewer than 10 prompts in total (19 participants) to exclude users who did not sufficiently use the LLM.
    \item Removing superusers (3 participants). These users are considered statistical outliers according to the \textit{two-standard-deviation criterion}, as they considerably exceed the mean number of 70 prompts per person (59 after removal); they sent between 235 and 415 prompts.
\end{enumerate}

This left us with the data of 64 participants from 25 project teams. 
Per treatment, we had 9 project teams in the control group, 8 in the context group, and 7 in the proactive group.
We give an overview of our individual participant demographics in  \autoref{table:combined_summary}. 

\begin{table*}[ht]
    \centering
    \small
    \caption{
    Background of participants.
    Likert-scale responses are shown as bars, with the following values on the \textcolor{likertleft}{left} and \textcolor{likertleft}{right} respectively:
    \newline
    [D-A] \textcolor{likertleft}{\textit{Strongly disagree}} to \textcolor{likertright}{\textit{strongly agree}}\newline
    [N-D] \textcolor{likertleft}{\textit{Never}} to \textcolor{likertright}{\textit{Daily}}\newline
    [I-E] \textcolor{likertleft}{\textit{Very inexperienced}} to \textcolor{likertright}{\textit{Very experienced}}\newline
    [W-B] \textcolor{likertleft}{\textit{Much worse}} to \textcolor{likertright}{\textit{Much better}}}
    
    \begin{tabular}{lrlrlrl}
    \toprule
     & \multicolumn{2}{c}{\emph{Control}} & \multicolumn{2}{c}{\emph{Context}} & \multicolumn{2}{c}{\emph{Proactive}} \\
     & \multicolumn{2}{c}{(N=25)} & \multicolumn{2}{c}{(N=18)} & \multicolumn{2}{c}{(N=21)} \\
    \midrule
    \textbf{Background and Experience} & & & & & & \\
    Median Prog. Experience (yrs) & 3.0 & (n=23) & 2.0 & (n=15) & 3.0 & (n=18) \\
    Attended 'PE' Lecture & yes & (n=25) & yes & (n=18) & yes & (n=21) \\
    Perceived Prog. Experience & & & \\ \quad compared to Peers [W-B] & 
    \multicolumn{2}{l}{\raisebox{-.16\height}{\includegraphics[trim=0.1cm 0.66cm 0.11cm 0.36cm, clip=true, width=0.17\linewidth]{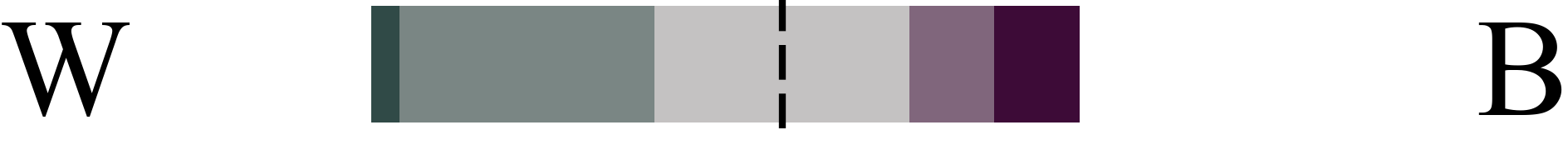}}} & 
    \multicolumn{2}{l}{\raisebox{-.16\height}{\includegraphics[trim=0.1cm 0.66cm 0.11cm 0.36cm, clip=true, width=0.17\linewidth]{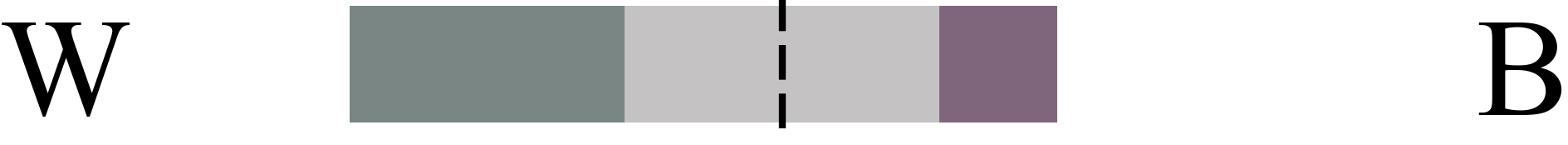}}} & 
    \multicolumn{2}{l}{\raisebox{-.16\height}{\includegraphics[trim=0.1cm 0.66cm 0.11cm 0.36cm, clip=true, width=0.17\linewidth]{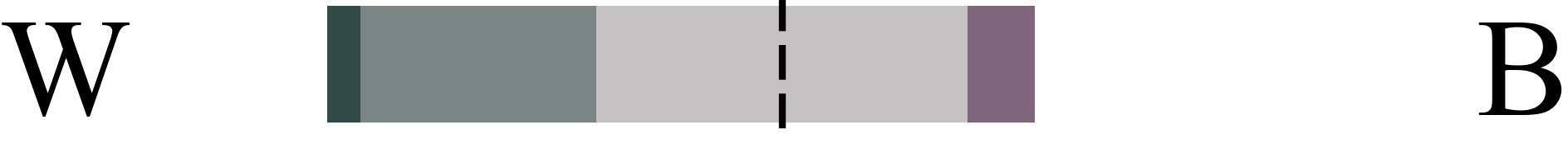}}} \\
    Logical Prog. Experience [I-E] & 
    \multicolumn{2}{l}{\raisebox{-.16\height}{\includegraphics[trim=0.1cm 0.66cm 0.11cm 0.36cm, clip=true, width=0.17\linewidth]{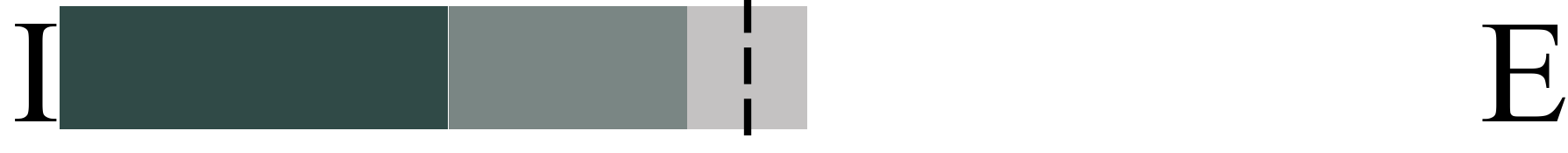}}} & 
    \multicolumn{2}{l}{\raisebox{-.16\height}{\includegraphics[trim=0.1cm 0.66cm 0.11cm 0.36cm, clip=true, width=0.17\linewidth]{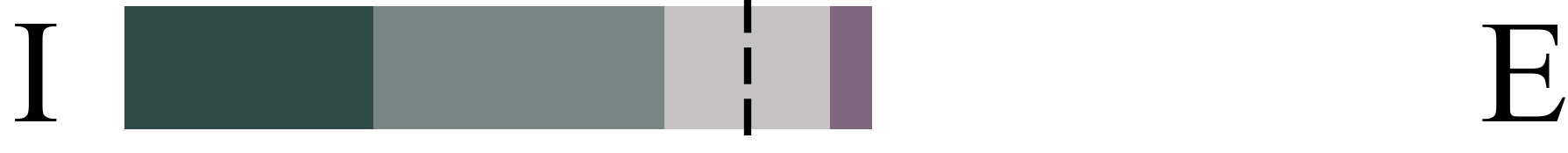}}} & 
    \multicolumn{2}{l}{\raisebox{-.16\height}{\includegraphics[trim=0.1cm 0.66cm 0.11cm 0.36cm, clip=true, width=0.17\linewidth]{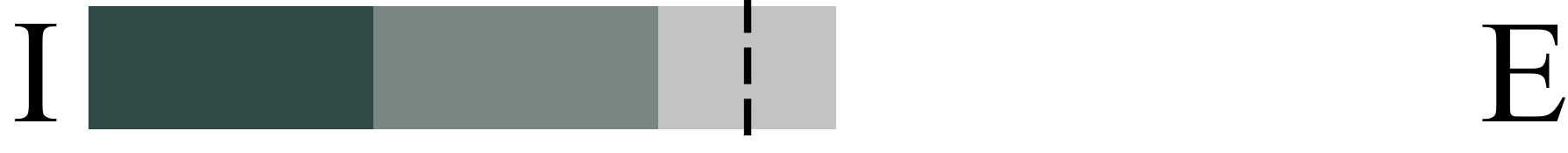}}} \\
    
    \midrule
    \textbf{Opinions on Chatbots} & & & & & & \\
    Use in daily life [N-D] & 
    \multicolumn{2}{l}{\raisebox{-.16\height}{\includegraphics[trim=0.1cm 0.66cm 0.11cm 0.36cm, clip=true, width=0.17\linewidth]{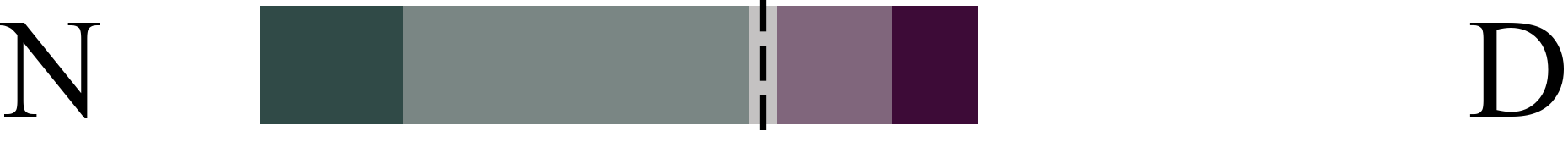}}} & 
    \multicolumn{2}{l}{\raisebox{-.16\height}{\includegraphics[trim=0.1cm 0.66cm 0.11cm 0.36cm, clip=true, width=0.17\linewidth]{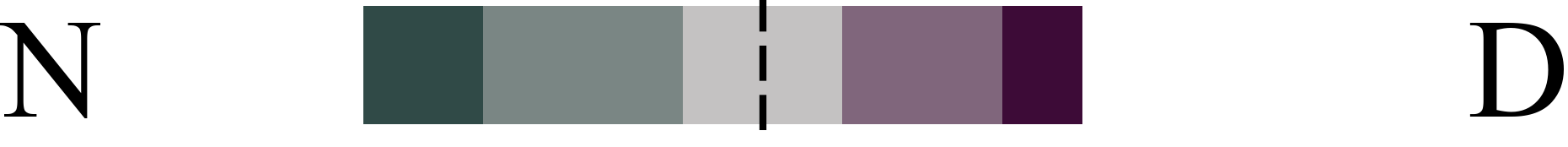}}} & 
    \multicolumn{2}{l}{\raisebox{-.16\height}{\includegraphics[trim=0.1cm 0.66cm 0.11cm 0.36cm, clip=true, width=0.17\linewidth]{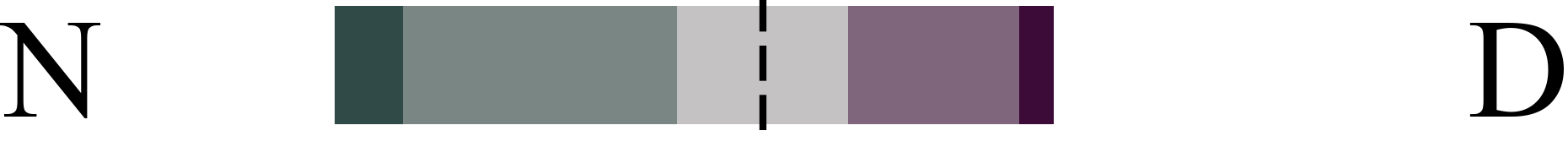}}} \\
    
    Chatbots are helpful... & & & & & & \\
    \hspace{1em}...in software dev. [D-A] & 
    \multicolumn{2}{l}{\raisebox{-.16\height}{\includegraphics[trim=0.1cm 0.66cm 0.11cm 0.36cm, clip=true, width=0.17\linewidth]{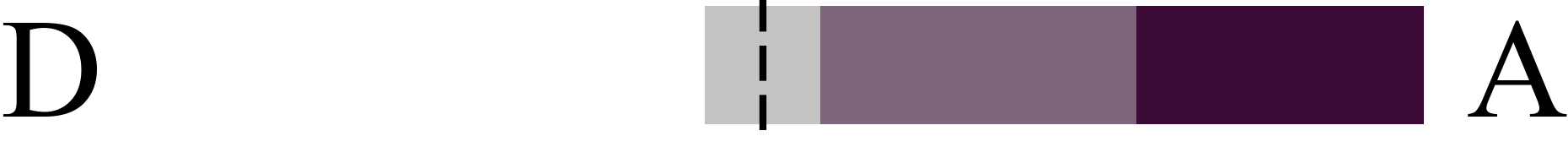}}} & 
    \multicolumn{2}{l}{\raisebox{-.16\height}{\includegraphics[trim=0.1cm 0.66cm 0.11cm 0.36cm, clip=true, width=0.17\linewidth]{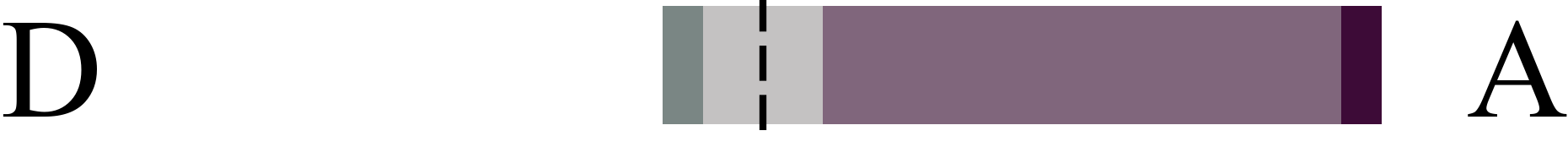}}} & 
    \multicolumn{2}{l}{\raisebox{-.16\height}{\includegraphics[trim=0.1cm 0.66cm 0.11cm 0.36cm, clip=true, width=0.17\linewidth]{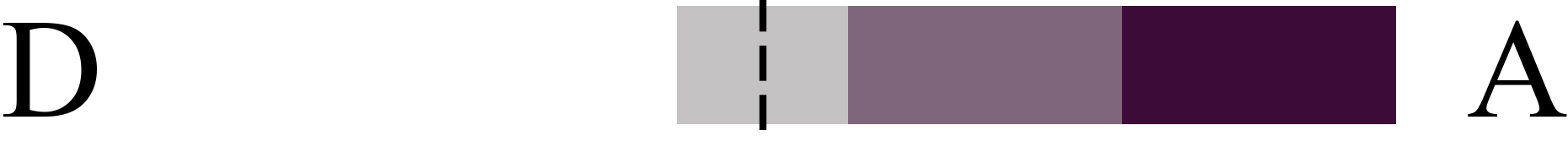}}} \\
    
    \hspace{1em}...in programming [D-A] & 
    \multicolumn{2}{l}{\raisebox{-.16\height}{\includegraphics[trim=0.1cm 0.66cm 0.11cm 0.36cm, clip=true, width=0.17\linewidth]{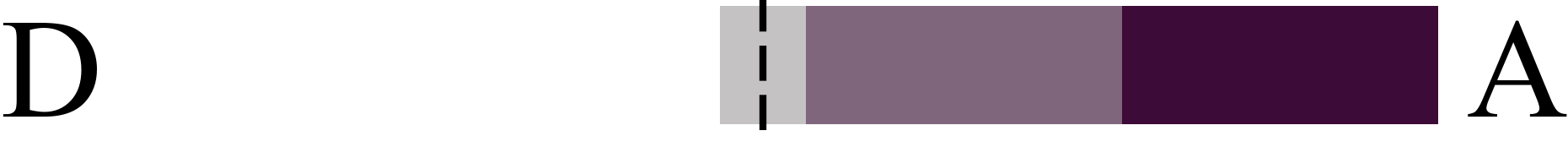}}} & 
    \multicolumn{2}{l}{\raisebox{-.16\height}{\includegraphics[trim=0.1cm 0.66cm 0.11cm 0.36cm, clip=true, width=0.17\linewidth]{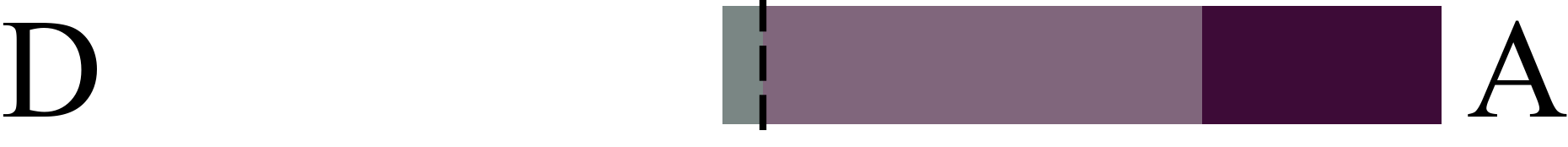}}} & 
    \multicolumn{2}{l}{\raisebox{-.16\height}{\includegraphics[trim=0.1cm 0.66cm 0.11cm 0.36cm, clip=true, width=0.17\linewidth]{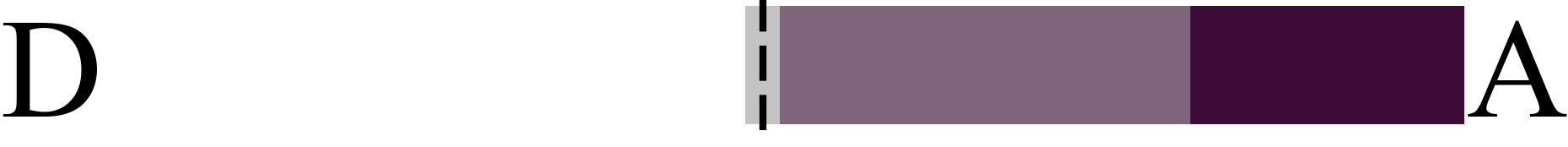}}} \\
    
    \hspace{1em}...in daily life [D-A] & 
    \multicolumn{2}{l}{\raisebox{-.16\height}{\includegraphics[trim=0.1cm 0.66cm 0.11cm 0.36cm, clip=true, width=0.17\linewidth]{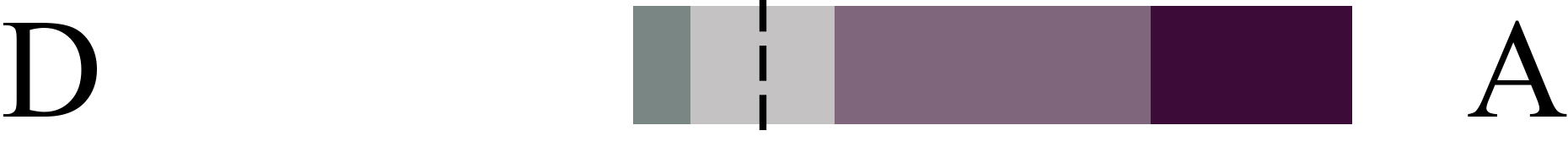}}} & 
    \multicolumn{2}{l}{\raisebox{-.16\height}{\includegraphics[trim=0.1cm 0.66cm 0.11cm 0.36cm, clip=true, width=0.17\linewidth]{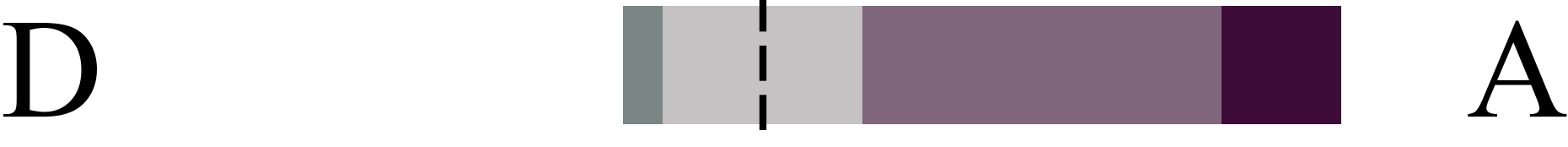}}} & 
    \multicolumn{2}{l}{\raisebox{-.16\height}{\includegraphics[trim=0.1cm 0.66cm 0.11cm 0.36cm, clip=true, width=0.17\linewidth]{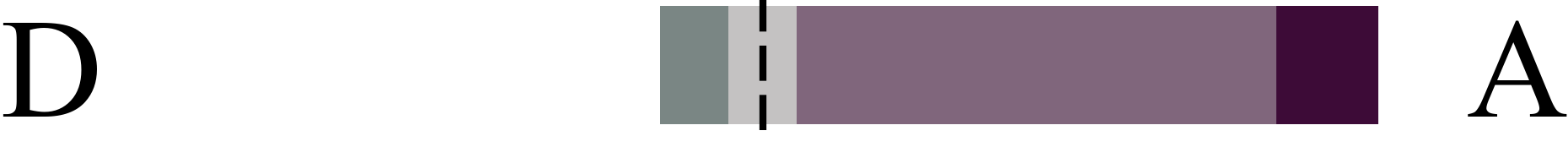}}} \\
    
    \hspace{1em}...in studies [D-A] & 
    \multicolumn{2}{l}{\raisebox{-.16\height}{\includegraphics[trim=0.1cm 0.66cm 0.11cm 0.36cm, clip=true, width=0.17\linewidth]{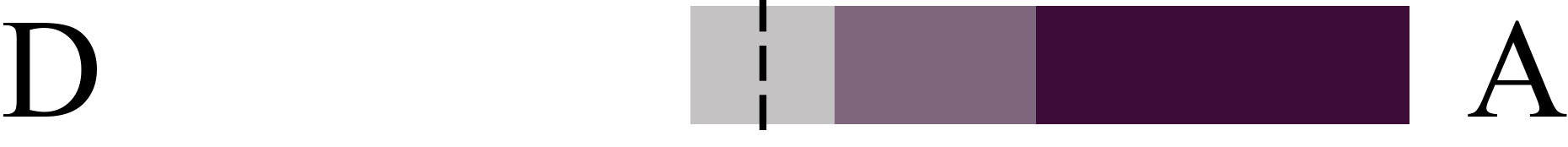}}} & 
    \multicolumn{2}{l}{\raisebox{-.16\height}{\includegraphics[trim=0.1cm 0.66cm 0.11cm 0.36cm, clip=true, width=0.17\linewidth]{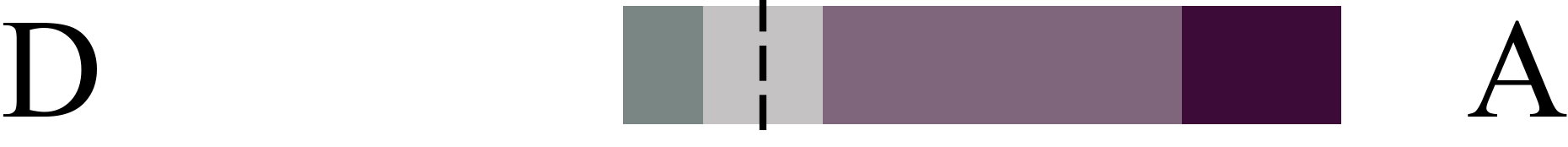}}} & 
    \multicolumn{2}{l}{\raisebox{-.16\height}{\includegraphics[trim=0.1cm 0.66cm 0.11cm 0.36cm, clip=true, width=0.17\linewidth]{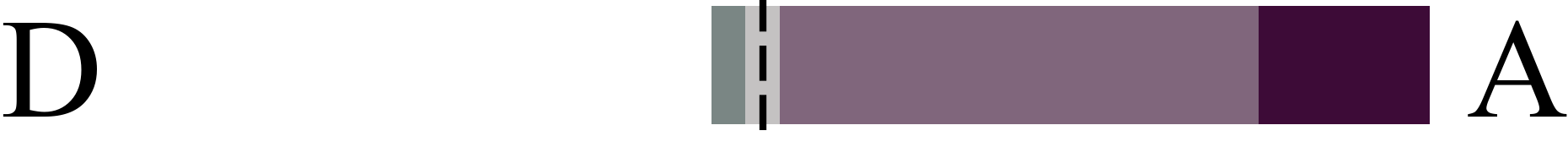}}} \\
    
    Chatbots are fun [D-A] & 
    \multicolumn{2}{l}{\raisebox{-.16\height}{\includegraphics[trim=0.1cm 0.66cm 0.11cm 0.36cm, clip=true, width=0.17\linewidth]{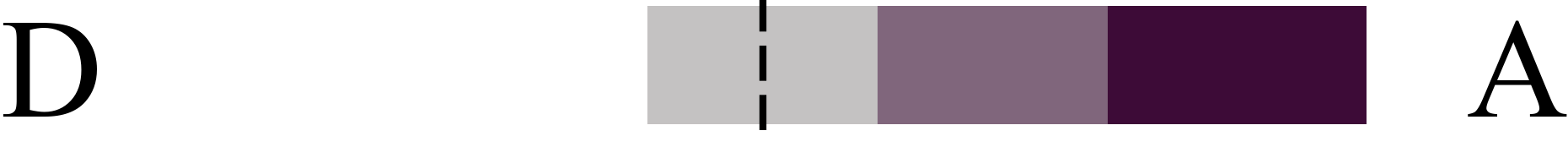}}} & 
    \multicolumn{2}{l}{\raisebox{-.16\height}{\includegraphics[trim=0.1cm 0.66cm 0.11cm 0.36cm, clip=true, width=0.17\linewidth]{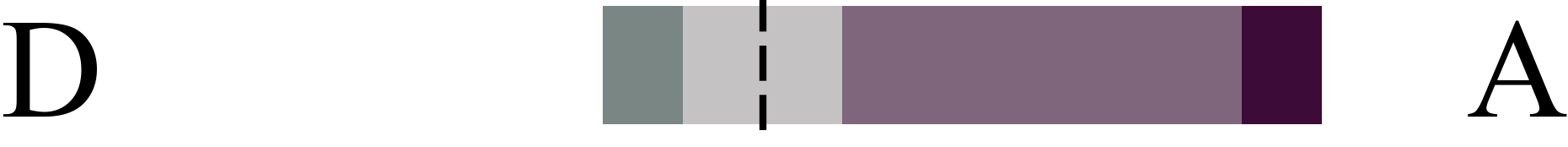}}} & 
    \multicolumn{2}{l}{\raisebox{-.16\height}{\includegraphics[trim=0.1cm 0.66cm 0.11cm 0.36cm, clip=true, width=0.17\linewidth]{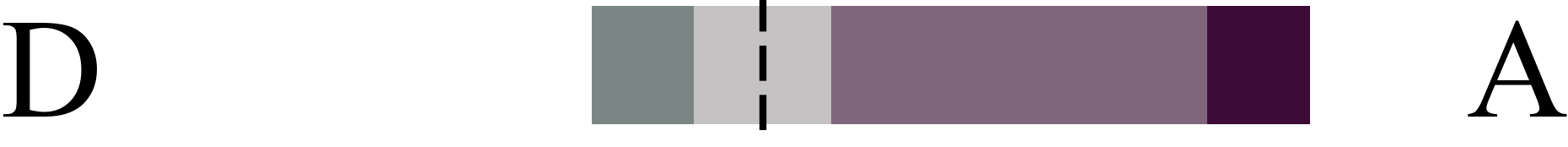}}} \\
    
    Confident using chatbots [D-A] & 
    \multicolumn{2}{l}{\raisebox{-.16\height}{\includegraphics[trim=0.1cm 0.66cm 0.11cm 0.36cm, clip=true, width=0.17\linewidth]{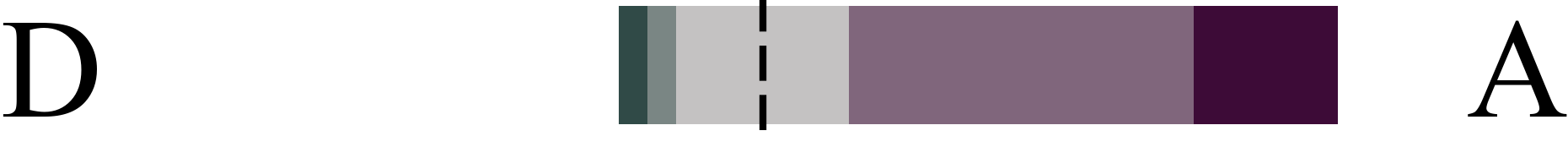}}} & 
    \multicolumn{2}{l}{\raisebox{-.16\height}{\includegraphics[trim=0.1cm 0.66cm 0.11cm 0.36cm, clip=true, width=0.17\linewidth]{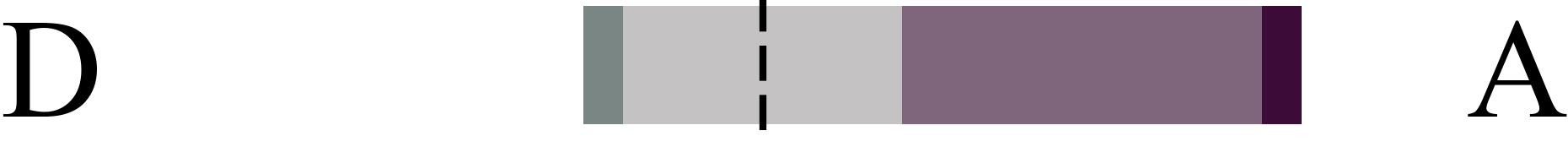}}} & 
    \multicolumn{2}{l}{\raisebox{-.16\height}{\includegraphics[trim=0.1cm 0.66cm 0.11cm 0.36cm, clip=true, width=0.17\linewidth]{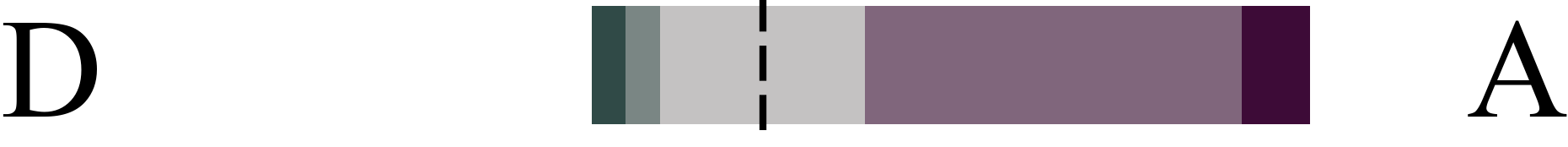}}} \\
    
    More future chatbot usage [D-A] & 
    \multicolumn{2}{l}{\raisebox{-.16\height}{\includegraphics[trim=0.1cm 0.66cm 0.11cm 0.36cm, clip=true, width=0.17\linewidth]{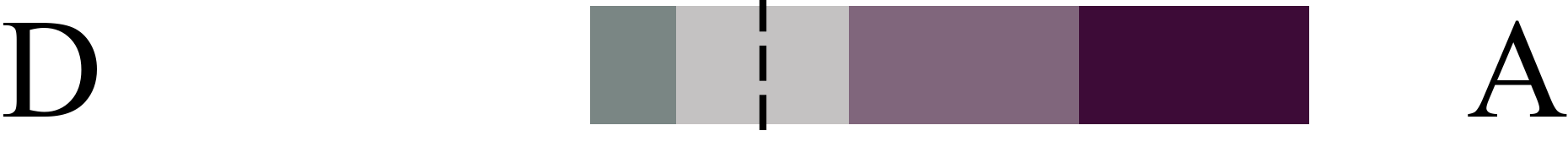}}} & 
    \multicolumn{2}{l}{\raisebox{-.16\height}{\includegraphics[trim=0.1cm 0.66cm 0.11cm 0.36cm, clip=true, width=0.17\linewidth]{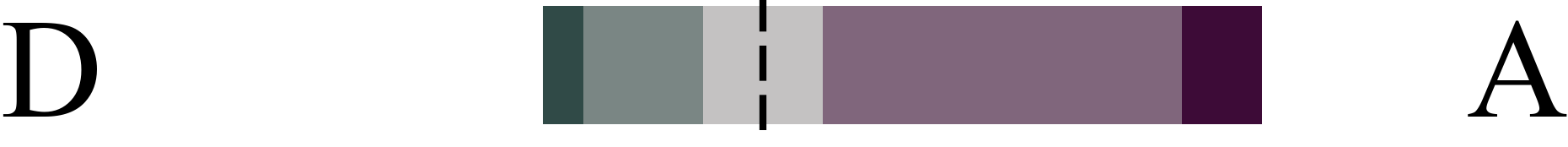}}} & 
    \multicolumn{2}{l}{\raisebox{-.16\height}{\includegraphics[trim=0.1cm 0.66cm 0.11cm 0.36cm, clip=true, width=0.17\linewidth]{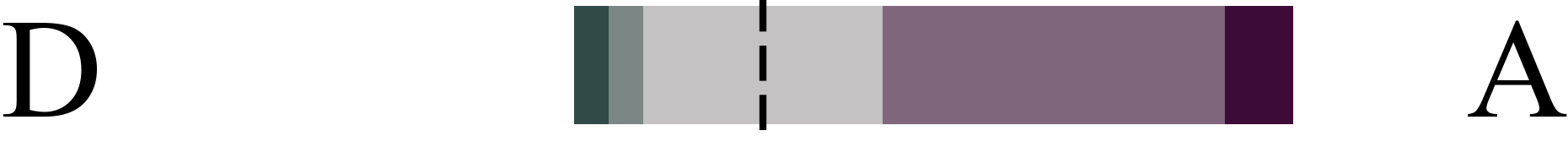}}} \\
    
    \bottomrule
    \end{tabular}
    \label{table:combined_summary}
\end{table*}

The programming experience level (\emph{Median Prog. Experience} in \autoref{table:combined_summary}) is comparable for all groups (confirmed by a Kruskal-Wallis-Test W$ = 2.84$, $p=0.24$), so different levels of programming experience cannot explain differences between groups in the dependent variables (further confounding variables, such as preceding programming courses, are also similar among the groups, see project's replication package). The same counts for the opinions on and usage of LLMs; although there are slight variations between the groups, there is no systematic difference, so that we can assume that the attitude toward LLMs is comparable between groups.

For the second research question, which focuses on perception rather than behavior,  we did not remove the outliers from the questionnaire responses, because super-users' opinions are still relevant here and do not skew any statistics. 
Instead, we removed responses only from participants with less than five prompts in the month previous to the questionnaire. 
For example, to be included in the data from the 2024-01-14 (second measurement date), users must have sent a minimum of 5 requests between 2023-12-06 (first measurement date) and 2024-01-14.
Additionally, if participants did not use the LLM in-between two measurement dates at all, they could not answer the questionnaire and thus are not included either.
This way, we ensure that participants have used the LLMs sufficiently to be able to give an informed response, while also preventing the loss of responses from participants who worked on the projects infrequently or became too frustrated to continue participating.

\section{Results and Discussion}
We present the results of each research question in this section, separated into data presentation and the discussion of the interpretation and implications of the results.

\subsection{\ques{1}: How does guiding a conversation influence developer behavior when interacting with an LLM?}
\subsubsection{User Behavior}
\paragraph{Conversation Length.}

\begin{figure}[ht]
    \centering
        \includegraphics[width=1\textwidth]{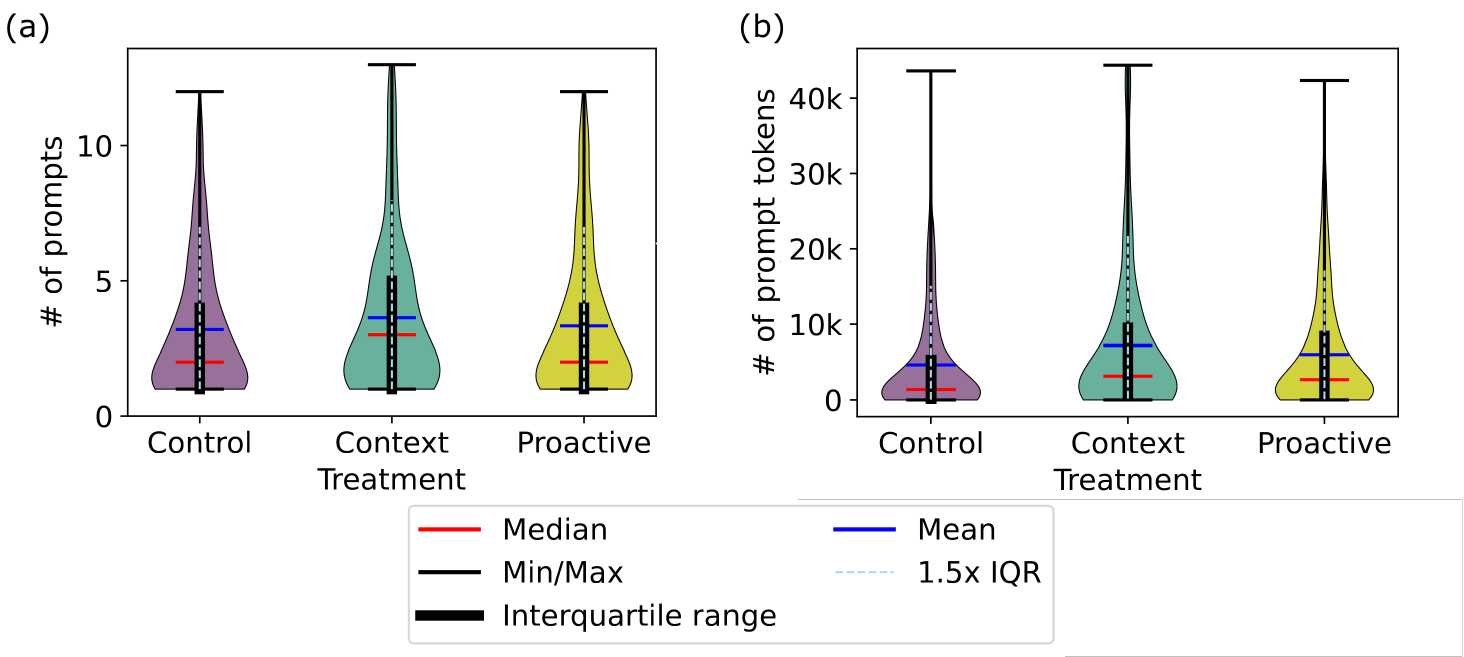}
    \caption{Average conversation length by treatment group per conversation by number of prompts per conversation (a) and number of prompt tokens per conversation (b)}
    \label{fig:treatment_summary_detailed}
\end{figure}

After removing outliers (conversations where the amount of prompts deviated more than two standard deviations of the mean), participants wrote on average 3.1 prompts per conversation in the \control{}, 3.6 in the \context{}, and 3.3 in the \proactive{} group.
For the number of tokens per prompt per conversation, also with outliers removed, there are, on average, 4638 for the \control{}, 7191 for the \context{}, and 5995 for the \proactive{} group. 
We present an overview in \autoref{fig:treatment_summary_detailed}.

A Kruskal-Wallis test shows no significant difference in the number of prompts (W$ = 3.39$, $p= 0.182$, Cliff's delta ranging from -0.07 to 0.06), while it does for the prompt tokens (W$ = 17.18$, $p < 0.05$, Cliff's delta ranging from -0.21 to 0.04). 
A Post-hoc Dunn's Test with Bonferroni correction confirmed the differences to be significant between the \control{} and \context{} ($p<0.05$, Cliff’s $\delta=-0.21$), and the \control{} and \proactive{} group ($p<0.05$, Cliff’s $\delta=-0.17$). Notably, these differences are not merely significant based on the large number of tokens, but also the effect sizes are larger than for the number of prompts.

\begin{figure}
    \centering
    \includegraphics[width=\columnwidth]{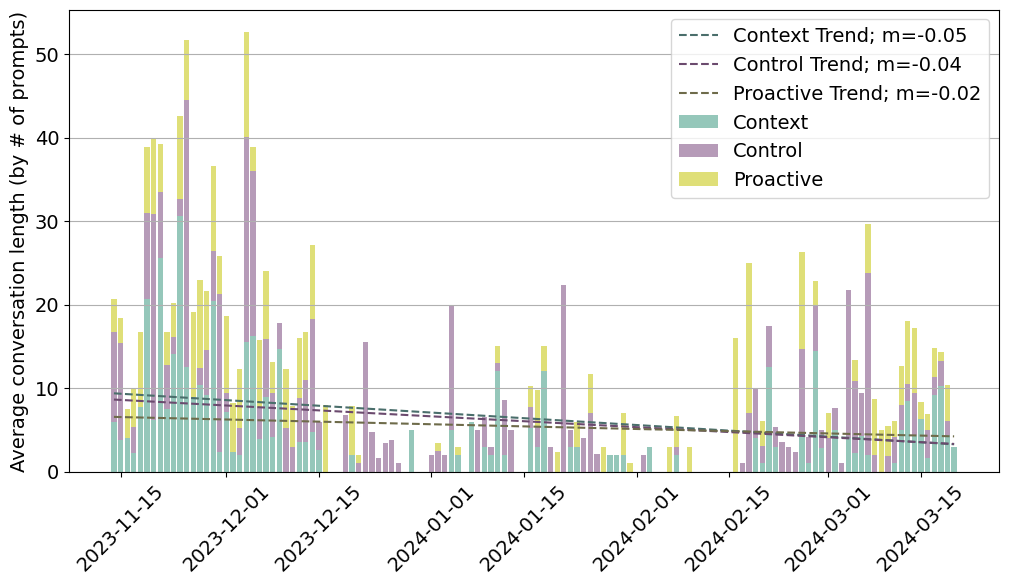}
    \caption{Length of conversations over time, averaged per day and divided by treatment}
    \label{fig:converstaion_length_over_time}
\end{figure}

\autoref{fig:converstaion_length_over_time} shows that the number of prompts per conversation drops with the project duration. This holds for all groups and is most severe for \context{} treatment. Interestingly, \proactive{} treatment seems already right at the beginning requiring fewer prompts. 

\begin{figure}[ht]
    \centering
    \includegraphics[width=\linewidth]{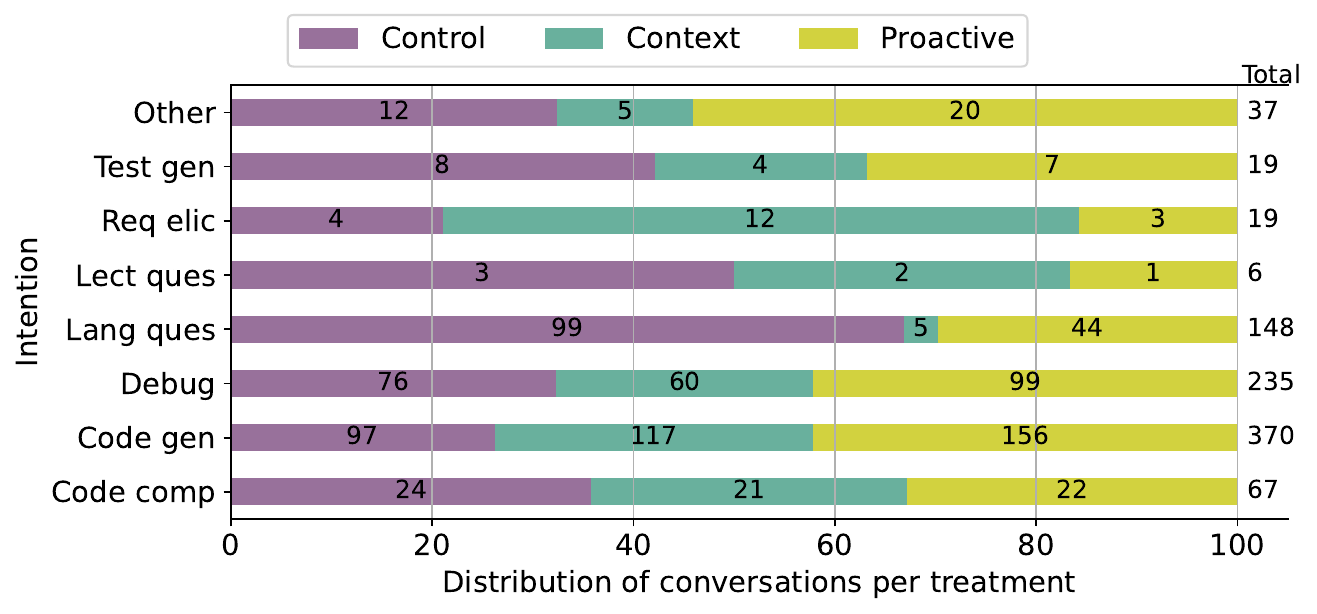}
    \caption{Conversation intentions per treatment. The number of conversations with an intention is shown in the bars; the total number of conversations is shown on the right}
    \label{fig:percentualconditioninintention}
\end{figure}

\paragraph{Specified Intentions.}
\label{sec:intentions}
\autoref{fig:percentualconditioninintention} gives an overview of the distribution of conversation intentions. 
Across all groups, the most frequent intentions are \textit{code generation} (370), \textit{debugging} (235), and \textit{language questions} (148). Notably, participants in the \emph{control} group started far more conversations with the intention \textit{language question} (\textit{Lang ques}) than any other group ($66.9\,\%$ of such  conversations, compared to 3\,\% (\context{}), and 29\,\% (\proactive{})). It is their most frequent intention (99 of 323 conversations, 30\,\% of all their conversations).
In the \context{} and \proactive{} group, the most frequent intention was \textit{code generation} (\textit{Code gen}) with 117 of 226 (54\,\%) and 156 of 352 (44\,\%) conversations, respectively.
These differences in frequency are statistically significant ($\chi^2 = 111.40 , df = 14, p < 0.001$).
Furthermore, the \proactive{} group used code generation intentions 60\,\% and 33\,\% more often than the \control{} and \context{} groups, respectively. 

\begin{figure}[ht]
    \centering
    \includegraphics[width=1.1\columnwidth]{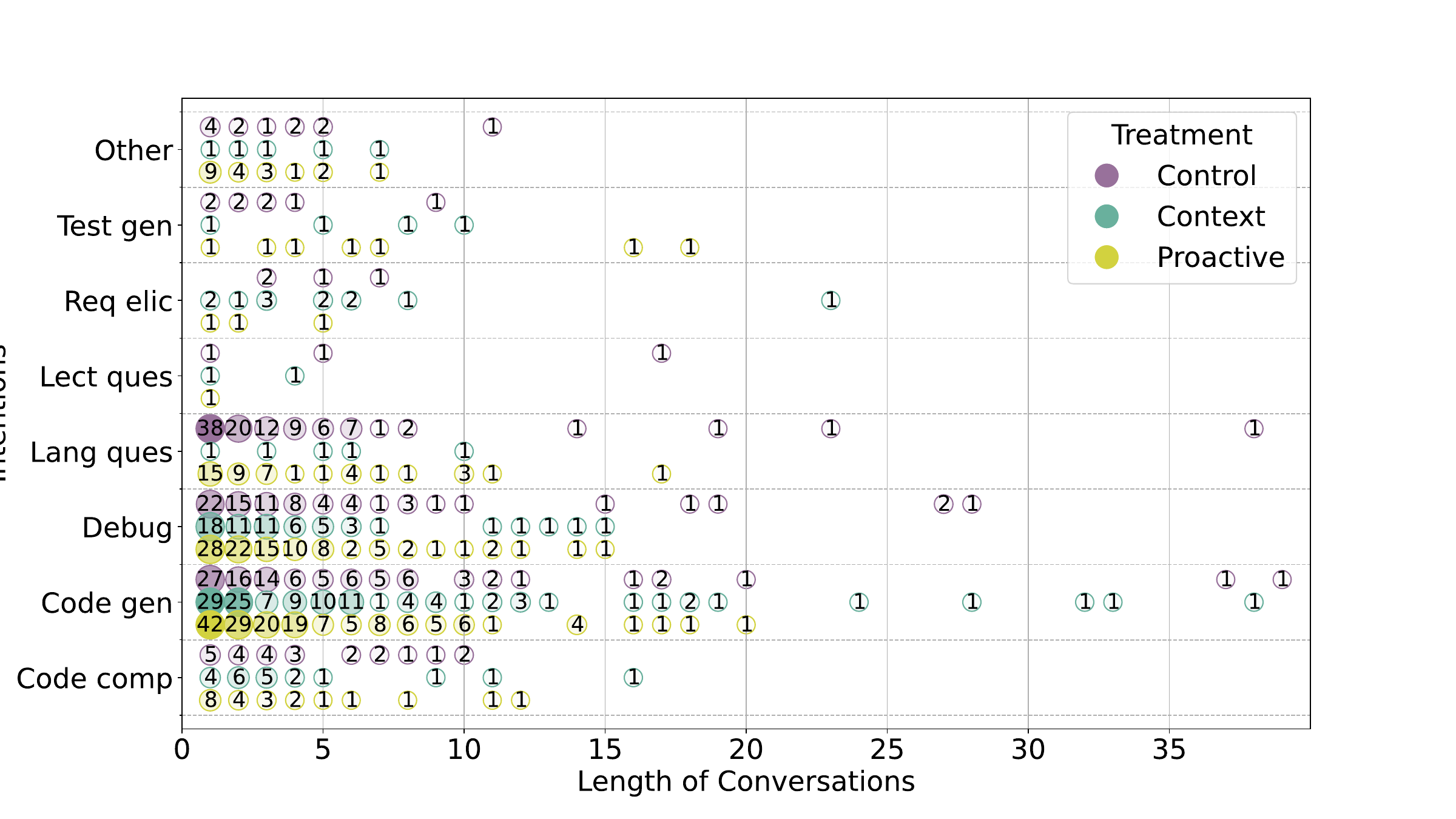}
    \caption{Distribution of conversation lengths per intention 
    }
    \label{fig:conversationlengthintention}
\end{figure}

\paragraph{Interaction of conversation length and specified intentions.} We also evaluated whether conversation length in terms of number of prompts is specific to the conversations' intention. Most of the conversations are kept short, but intentions \textit{language questions}, \textit{debugging}, and \textit{code generation} relate to longer conversations. 
Figure \ref{fig:conversationlengthintention} shows how many conversations (numbers in circles) with a specific conversation length (x-axis) were found for each intention (y-axis). The median length of conversations varies between 2 (custom intentions, marked as \textit{other}) and 4 (\textit{test generation}).

A two-way ANOVA indicates a potentially significant impact of the intention on the conversation length (F $=2.02$, df$ = 7$, $p = 0.049$), but shows no significant impact of neither the treatment (F $= 2.3$, df$ = 2$, $p > 0.05$) nor the interaction of treatment and intention (F $= 0.8$, df$ = 14$, $p > 0.05$). In post-hoc tests, the significant effect of intention vanishes.

\paragraph{Intention alignment with conversation content.}
\label{sec:intentionalignment}

The manual analysis of 135 conversations showed whether and how participants' prompt contents diverge from their initially specified intentions.

\autoref{table:treatment_summary_mismatches} provides details on misalignments between intention and actual conversation.
Over all groups, $38.52$\,\% of the analyzed conversations contain only prompts that entirely match the user-specified intention. Across all groups, about 25\,\% of conversations have been started with a prompt that does not match the user-specified intention (the upper part of the table, column(\(\%_{1st}\))), which is similar for all treatments. If any deviation occurs, it usually occurs with the second prompt (columns \(\text{med}_{pos}\) and \(\text{avg}_{pos}\)) for all treatments. Furthermore, 35.56\,\% of all conversations in the \control{} group, 37.78\,\% in the \context{} group, and 26.67\,\% in the \proactive{} group had more than half of their prompts misaligned with the user-specified intention. Thus, a considerable number of conversations do not align with the user-specified intention and shift quickly, mostly after the first answer. Interestingly, the \proactive{} group has the highest agreement in intention and actual conversation.

Breaking down the conversations per group, 37.78\,\% of conversations in the \control{} group, 31.11\,\% in the \context{} group, and 46.67\,\% conversations in the \proactive{} group solely contain aligned prompts. A $\chi^2$-test of independence on the number of aligned and misaligned prompts shows no significant difference between the alignment of prompts and the treatments ($\chi^2 = 5.31 , df = 2, p = 0.069$).

While these misalignments may have influenced previously analyzed variables for this RQ, the majority of conversations and prompts remain aligned; thus, we consider the data suitable for analysis of previous RQs. The inability to express one's intentions and the tendency to shift intentions quickly during a conversation is interesting in its own right. This insight has not been empirically observed in related studies, yet it will affect future chat mechanisms.

\begin{table}
    \centering
    \footnotesize
    \caption{Misalignments between user-specified intentions and prompt contents, split by treatment group or intention. 
    Columns describe conversations with at least one misaligned prompt (\(\%_{min1}\)), more than half misaligned prompts (\(\%_{half}\)), and the mean and median first positions in a conversation where a misalignment occurs (\(\text{avg}_{pos}\), \(\text{med}_{pos}\)). Prompt-level metrics include the percentage of first prompts of conversations that were misaligned (\(\%_{1st}\)) and the percentage of prompts where an intention that was assigned by us did not align with the user’s given intention (\(\%_{FN}\)).  }

        \begin{tabular}{llccccccc}
        \toprule
         \multicolumn{2}{l}{Prompt-level} & & $\%_{1st}$ &  & &  &  & $\%_{FN}$ \\
         \multicolumn{2}{l}{Conversation-level}  & N & & $\%_{min1}$ & $\%_{half}$ & $\text{avg}_{pos}$ &  $\text{med}_{pos}$ & \\
        \midrule
        \midrule
               \multirow{3}{*}{\begin{sideways}\begin{minipage}[t]{0.8cm}
              \centering {\scriptsize{By treatment}} \end{minipage}\end{sideways}} 
        & Control   & 45 & 26.67 & 62.22 & 35.56 & 2.46 & 2.0 &\\
        & Context   & 45 & 28.89 & 68.89 & 37.78  & 2.19 & 2.0 &\\
        & Proactive & 45 & 22.22 & 53.33 & 26.67  & 2.00 & 2.0 &\\
        \midrule
               \multirow{5}{*}{\begin{sideways}\begin{minipage}[t]{1.3cm}
              \centering {\scriptsize{By given intention}} \end{minipage}\end{sideways}} 
        & Code gen        & 52 & 36.54 & 75.00 & 46.15 & 2.03 & 2.0 & 11.43 \\
        & Code comp     & 7  & 14.29 & 85.71 & 42.86 & 2.00 & 2.0 & 5.71 \\
        & Debug              & 37 & 10.81 & 48.65 & 10.81 & 2.72 & 3.0 & 22.86 \\
        & Lang ques      & 28 & 14.29 & 46.43 & 25.00 & 2.92 & 2.0 & 45.71 \\
        & Test gen        & 3  & 66.67 & 66.67 & 66.67 & 1.00 & 1.0 & 2.86 \\
        \bottomrule
    \end{tabular}
    % }
    \label{table:treatment_summary_mismatches}

\end{table}

Viewing the data by user-specified intention (lower part of \autoref{table:treatment_summary_mismatches}), participants indicating that they intend to debug deviate the least. Additionally, if participants deviate from \textit{debugging}, they do so later in the conversations compared to any other intention (on median after the 3rd prompt; column `Median$_{pos}$'). 
A potential reason for this observation is that participants might need more prompts to find a bug using the LLM than they need to fulfill other tasks. 
The highest number of deviations occurs when participants specify to generate code and tests (85.71 and 66.67\,\%, column $\%_{min1}$). In cases such misalignments occur, the most often used intention is \textit{language questions} (45\,\%; column $\%_{FN}$), which might mean that, once code is returned by the LLM, it needs to be understood, so the intention shifts to \textit{language questions} or \textit{code comprehension}.

\begin{mybox}{\ques{1}: Results}
Conversational behavior of participants shows no major differences among treatments regarding the number of prompts, but regarding the number of prompt tokens per conversation. Conversations also get shorter over time whereas \proactive{} already starts with short conversations.

There is a significant imbalance of our treatment groups between the number of conversations for specific intentions, despite the overall number of conversations being similar. This difference is most prominent for participants asking about language-specific details. In the \control{} group, participants are starting conversations with this intention almost 20 times more often than participants in the \context{} group. Moreover, programming related intentions, such as code generation and debugging intentions, are used substantially more often for the \proactive{} group than for the two others. 

About a quarter of conversations start with a mismatch of the stated intention and the actual content of the prompt. This number rises after the first answer and is most pronounced for \emph{code generation} and \emph{comprehension}. The \proactive{} group has the highest alignment of actual and user-stated intentions.

\end{mybox}

\subsubsection{Discussion: Effects of guidance levels on developer-LLM-interactions}
The results regarding specified intentions align with other studies, indicating that developers use AI assistants for code generation and debugging\cite{surveyonusage,mailach,copilotcasestudy,codegendebugging}. 
While not entirely surprisingly, it is interesting that the participants of our study only rarely asked the LLM for test generation, while Sergeyuk et al.\cite{surveyonusage} found that this was an often used intention. This discrepancy might stem from the different contexts of our programming-beginner setting, compared to the professional setting of their study, as automated tests were not mandatory for the programming project (yet the final product must still run and fulfill the requirements).

Looking at the specified intentions by treatment, the \control{} group specifies most often language questions. This intention describes participants wanting to start a conversation to learn more about a programming language. 
At first sight, this might indicate that the control setting motivates participants toward using the LLM to dive more into understanding the programming language, while the \proactive{} and \context{} nudges participants toward using the LLM to provide a working software product, less toward learning something about the code. 
However, in conjunction with the result that the actual content of the conversation in many cases does not align with the specified intention draws a different picture. Especially conversations with the specified intention of \emph{code generation} often contain prompts about language-specific details (19.5\,\% of prompts within code generation conversations). 

Thus, even though the LLM interface explicitly guides users to think about what they want from the conversation, they have trouble doing so. We interpret this to mean that early-career developers, like those in our sample, might struggle with deciding up-front what exactly they want from the LLM. If they are not sure about their intention, it is likely that their prompt reflects that in vagueness, such that the LLM is used ineffectively. This insight shows the need to clarify intentions of users upfront, which may also yield a more effective prompt design. We are not aware of any user study that identified the extend of this mismatch despite its importance for the soundness of empirical evalutions and design of new chat mechanisms.

Regarding efficiency, we see a longitudinal trend towards shorter discussions, indicating that developers get more efficient in their LLM communication over time as they require fewer prompts in a single discussion. A main observation is that with the \proactive{} treatment, conversations were already short at the first time period, which stands out and may point to an increased efficiency.

In our specific setting, intentions within a conversation could not be changed without starting a new conversation; a trade-off of our study design. This can be unproductive, as a new conversation does not have the relevant context. We found clear evidence that a static intention context is insufficient, as the natural flow of conversations shifts the focus. The \proactive{} treatment accounts for this dynamic nature and shows the lowest misalignments in intentions and actual conversation content. A possible explanation is also that this treatment ensures that the user stays on track with its initial intent and does not slide into other intention activities.
So either requiring users to specify their intentions with every prompt or leveraging systems, that predict intentions without additional user input, could improve effectiveness of LLMs. 

\begin{samepage}
\begin{mybox}{}
    The answer to \ques{1} is nuanced. 
    At first glance, there is no significant difference in developers' general behavior when interacting with the LLM, regardless of the treatment. Even for the \context{} group, we observed no substantial behavior changes, despite identifying frequent misalignments of intentions throughout conversations. 
    A closer look reveals that the \proactive{} group 
    seems to better align their prompts to their initial conversation intention. However this difference is not significant. Additionally, participants in this group used the LLM most often for development specific intentions, such as debugging and code generation, and used fewer prompts to finish a conversation, pointing to higher efficiency.
    Nevertheless, this could show a tendency of the ability of dynamic approaches to support participants in keeping their conversations aligned with their initial goal when using the LLM specifically for software development.
\end{mybox} 
\end{samepage}

\subsection{\ques{2}: How do developers perceive LLMs with guided conversations?}
\label{sec:req2}

\subsubsection{Developer Perception}

\begin{figure*}
    \centering
    \includegraphics[width=\linewidth]{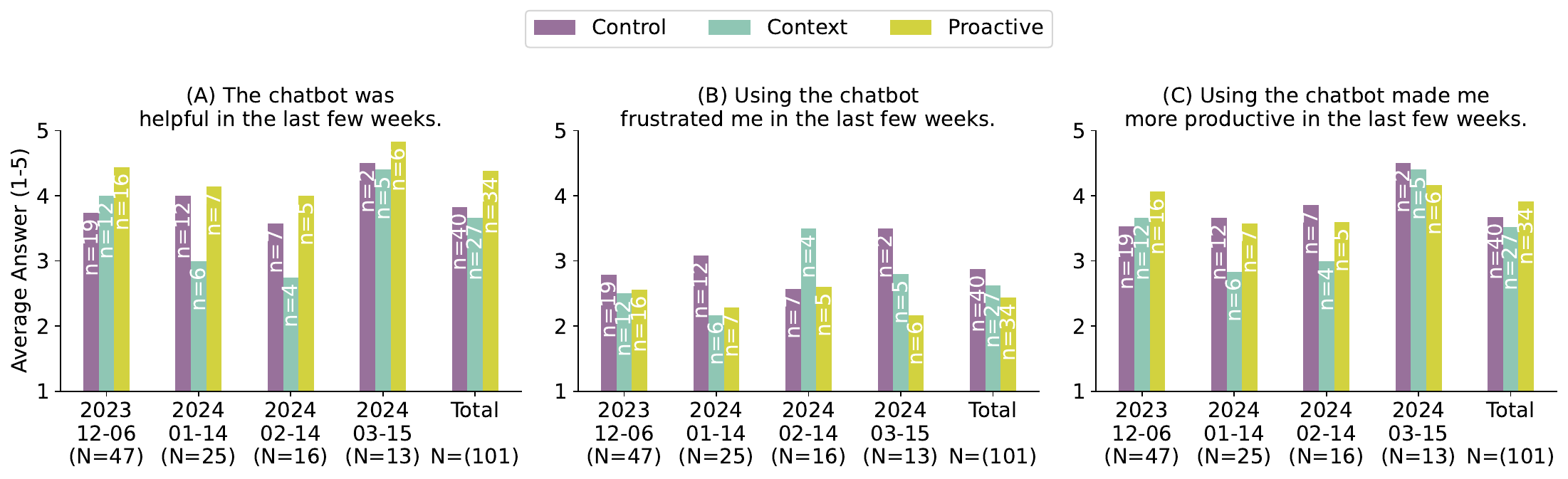}
    \caption{Averaged answers for perceived helpfulness, frustration, and productivity. The answers range from 1 to 5, with 1 being ``strongly disagree'' and 5 ``strongly agree''. Higher values indicate a positive rating in A and C, but negative in B.}
    \label{fig:questionnaireresults}
\end{figure*}

To evaluate \ques{2}, we show the monthly survey results of participants in \autoref{fig:questionnaireresults}.
As is common in longitudinal studies (especially in voluntary university experiments), the number of participants reduces with each month.
We started with 47 replies for the first survey and ended up with only 13 replies for the last.
Starting with the second survey, the group sizes were small, making significance tests underpowered and the results likely inconclusive.  
Therefore, we compared the results of surveys 2–4 without testing for significance.  
This approach helps identify tendencies between treatments that can inform future research directions.

The first survey was conducted during the period with the highest development effort for the projects, so more developers answered it. 
Therefore, we performed pairwise Mann-Whitney U-tests for this survey to compare individual groups. 
Thus, we compare the values for tendencies, but note that differences are minor.
More detailed information on this can be found on our Web site.

\paragraph{Helpfulness.}
The total average value on perceived helpfulness (\textit{Total} in \autoref{fig:questionnaireresults}) is 3.9 for the \control{}, 3.5 for the \context{}, and 4.3 in the \proactive{} group (5 meaning very helpful), so all groups found the LLM to be at least somewhat helpful.
A pair-wise Mann-Whitney U-test revealed significant differences between the \proactive{} and both, the \control{} and the \context{} group (U$=452.0$, $p=0.008$ and U$=269.0$, $p=0.003$ respectively).
Overall, there is a downward trend of perceived helpfulness for all groups from the first to the third survey, and an increase for the last survey. The \proactive{} group reported the highest helpfulness rating (\autoref{fig:questionnaireresults} A) in all surveys, while the \control{} group reported the lowest in the first survey (U = $88.5$, $p = 0.025$).
Cliff's delta effect sizes for the first survey range from $-0.41$ (when comparing the \control{} and \proactive{} group) to $0.06$ (\context{} to \control{}), confirming that the \proactive{} group viewed the LLM more helpful at this point.
The \context{} group has the lowest helpfulness rating in the second and third surveys, dropping notably from the first.

\paragraph{Frustration.} 
The average participant frustration scores are 2.9 for \control{}, 2.7 for \context{}, and 2.4 for \proactive{} (1 meaning not frustrated), indicating low frustration.
There does not seem to be an overall picture of how frustration evolves. Participants in the \control{} group showed the highest frustration (\autoref{fig:questionnaireresults} B) in the first two surveys, then align with the \proactive{} group on a medium-low level. 
After a considerable increase in frustration from the second to third survey, the \context{} group is the most frustrated in the third survey. 
The \proactive{} group consistently reported low frustration levels. None of the differences in the first survey is statistically significant, and the Cliff's delta ranges from $-0.14$ to $0.11$.

\paragraph{Productivity}
On average, the total perceived productivity of participants is 3.8 for the \control{} and \proactive{}, and 3.4 for the \context{} group  (5 meaning very productive).
Participants' perceived productivity (\autoref{fig:questionnaireresults} C) increased in each survey for the \control{} group, showing the highest ratings of all groups in the second and third surveys. 
Again, with only two people from the \control{} group giving responses, we cannot reliably compare their results for the last survey.
The \context{} and \proactive{} groups' perceived productivity drops from the first to the second survey.
However, the drop is more pronounced for the \context{} group, resulting in the lowest perceived productivity in the second and third surveys. Again, none of the differences in the first survey are statistically significant, with Cliff's delta effect sizes ranging from medium $-0.30$ (comparing the \context{} and \proactive{} group) to null $0.0$ (\context{} and \control{}).

\begin{mybox}{\ques{2}: Results}
    The \proactive{} group perceived the LLM as the most helpful and as more beneficial to their productivity while rarely feeling frustrated, while maintaining the highest user engagement throughout the experiment.
    The \context{} group viewed the LLM the most critical, while both, the \context{} and \control{} groups felt frustrated by the LLM at some point during the semester.
\end{mybox}

\subsubsection{Discussion: Developer perceptions of guided conversations with LLMs}

The perception of participants differs marginally between groups, indicating that all variants of LLMs are perceived as helpful, supporting productivity, and only rarely lead to frustration. Especially the \proactive{} group shows the most positive perception, and one participant highlighted this in the open text field \xquote{[...] I can only say that programming itself can be frustrating sometimes and I think without the chatbot[...] it would be even more exhausting and frustrating}{P1}.

The consistently lower helpfulness and productivity ratings of the \context{} group might be linked with the misalignment of the user-specified with the actual intentions of a conversation, which is most prominent for the \context{} group (cf.\ Section~\ref{sec:intentionalignment}). The intention is not used for the \control{} group, and has limited effect for the dynamic conversation in the \proactive{} group. So, they do not receive misaligned responses, resulting in a more positive perception.

In general, these results align with previous studies on perceived productivity\cite{littleimprovement}, and contradict others that found significant productivity improvements\cite{copilotcasestudy} or that observed increased frustration and no productivity gains\cite{surveyonproductivity}. This further highlights the mixed effects of LLMs. Thus, these findings add another piece to the puzzle to better understand how different prompting techniques and guidance mechanisms affect software developers.

For the \proactive{} group, we further manually analyzed how users adopted the features of follow-up questions and improved prompts. About two-thirds of participants used follow-up questions or improved prompts, indicating that the functionality holds some appeal for the majority of users. Looking more closely, especially for the user-specified intentions of \emph{code generation} and \emph{language questions}, participants used either follow-up questions or improved prompts to continue the conversation. Over time, this decreased, which could indicate that participants might have learned how to improve their conversations, thus not needing any more assistance. Furthermore, since there does not seem to be negative perception of participants regarding guidance, this might be a promising endeavor to integrate relevant tips and techniques in custom-based chatbots, at least for early-career developers.

\begin{mybox}{}
    
    To answer \ques{2}, we find that LLMs are perceived differently depending on the treatment. 
    Especially developers using the \proactive{} LLM perceived the LLM the most positive, suggesting that active guidance features positively impact developers perceptions of LLMs. This is noteworthy, considering that we used a conservative, low-impact approach.
    By contrast, developers using the \context{} LLM rated it the least helpful,  which is likely due to the misalignment of given prompts with conversational contents. 
    These findings suggest that, while guiding LLMs can enhance user experience, not all treatments are equally well-received. 
    Given the better results in the \proactive{} group, enhancing developer-LLM-interactions with proactive guidance elements is a promising direction for future work.
\end{mybox}

\subsection{\ques{3}: How do different guidance and support mechanisms influence code produced by developers?}

\subsubsection{Results}
To answer \ques{3}, we evaluate the amount of generated code that is pushed into the repository, the modification rate, and the duration of how long generated code remains in the project.

\paragraph{Amount of generated code pushed to repositories.}
Over all treatments, participants made a median of 59 commits. Per treatment, the median commits by person were 72 (\control{}), 42 (\context{}) and 44 (\proactive{}) commits. 

\begin{wrapfigure}[12]{r}{0.38\textwidth}
    \vspace{-2em}
    \centering
        \includegraphics[width=\linewidth]{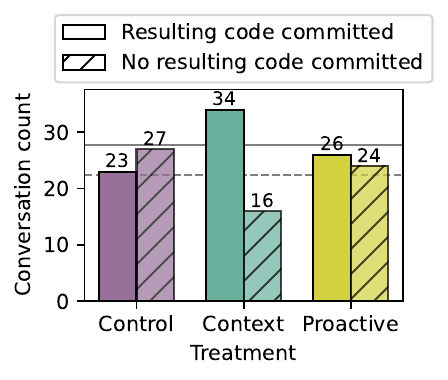}
    \vspace{-1em}
    \caption{Pushed vs.\ not pushed generated code per treatment; solid and dashed lines mark their respective averages.}
    
    \label{fig:codecommits}
\end{wrapfigure}
To verify whether participants used code generated by the LLM, we sampled 150 conversations for manual review (cf. Section~\ref{sec:rq}).
\autoref{fig:codecommits} shows that, for 46\,\% (23) conversations in the \control{} group, 68\,\% (34) in the \context{} group, and 52\,\% (26) in the \proactive{} group, generated code was committed to the repository. Thus, most generated code was committed in the \context{} group. This difference is significant between the \control{} group and \context{} group ($\chi^2 = 4.07$, df $=1$ , $p = 0.04$, Cliff's delta $= -0.22$), but it disappears after \textit{Bonferroni correction}.

\paragraph{Modification rate.}

Overall, participants change 74.01\,\% of the generated code over time.
\autoref{tab:commitschanged} gives an overview of the code changes by treatment ($\%_{\text{total}}$), showing that the \context{} and \proactive{} groups have a similar modification rate, and the \control{} group shows the lowest number of modifications. The difference of the total modification rates is significant ($\chi^2=20.95$, $p<0.01$).

We normalized the changes by computing the average percentage of modified lines per committed code segment ($\%_{\text{segment}}$) to better account for code-segment specific modifications (e.g., ignoring formatting related changes). A code segment comprises coherent code that was generated by the LLM, and pushed to the same file. 
The lowest modification rate still occurs in the \control{} group (60.25\%), but the rate of modification of the \context{} group is now more similar to the \control{} group, and the change for the \proactive{} group is negligible. After the normalization, the difference is not significant anymore, as shown by a Kruskal-Wallis test (W$ = 1.46$, $p= 0.48$).

\begin{table}[h!]
    \centering
    \caption{Changes made to the LLM-generated code after it was committed to the project, including 
    the total percentage of changes to generated code ($\%_{\text{total}}$), the average percentage of changes per committed code segment ($\%_{\text{segment}}$), and the average number of days until changes occurred ($\text{Time}_{\text{days}}$).}
    \begin{tabular}{lccccc}
        \toprule
        Treatment &
        $\%_{\text{total}}$ & $\%_{\text{segment}}$ & $\text{Time}_{\text{days}}$ \\
        \midrule
        \emph{Control} & 
        65.26 & 60.25 & 18.72 \\
        \emph{Context} & 
        76.47 & 65.54 & 8.70 \\
        \emph{Proactive} & 
        79.95 & 77.59 & 20.16 \\
        \bottomrule
    \end{tabular}
    \label{tab:commitschanged}
\end{table}

\paragraph{Duration until code is changed.}

Over all treatments, code remains unchanged for, on average, 15.37 days. 
Members of the \proactive{} treatment leave their code unchanged for the longest time period of 20 days, on average. 
The \control{} treatment follows with 18 days, while the \context{} treatment had the shortest time until changes occurred, averaging 8 days (right column, $\text{Time}_{\text{days}}$ in \autoref{tab:commitschanged}). This difference is not significant (W$=5.0$, $p=0.08$).

\begin{mybox}{\ques{3}: Results}
    Participants used more LLM-generated code in their projects if they used one of the guiding treatments. 
    The discrepancy is large with the \context{} group using generated code from 2 out of 3 conversations, compared to less than half of the conversations in the \control{} group.

    Around 70\,\% of LLM-generated code per file was eventually changed in all groups, with teams in the \proactive{} group changing code the most and the \control{} group changing the least. The alteration occurred the fastest in the \context{} group.
\end{mybox}

\subsubsection{Discussion: Effects of guidance levels on produced code}

The \context{} group had more conversations resulting in generated code being pushed. This is in line with the results of Spasić and Janković\cite{rolepromptinginse}, who observed that incorporating roles, specific instructions, and seed words into prompts improved model outputs to a more human-like level. Since the \context{} treatment works similarly, this might explain that the \context{} groups committed more generated code to the repository.
At first, it seems counterintuitive that the \context{} group has the highest number of generated code commits in the repositories, despite having lower helpfulness and productivity ratings. Simultaneously, the LLM code is modified the fastest, that is, after 10 days. A likely explanation is that, since the participants modify code more quickly, they are likely not satisfied with the generated code, but since they pushed it to the repository and now must modify it, they feel less productive and less supported by the LLM.
Interestingly, this insight contrasts another study regarding perceived productivity: Ziegler et al.\cite{ziegler} found that higher perceived productivity correlates with higher acceptance rate of generated code. This might also be caused by the different AI assistance (chat-based vs. direct IDE integration with code generation suggestions) or by the university context. The concrete underlying causes for this mismatch in perceived productivity and code contribution is an interesting research direction for future studies.

The overall high modification rates are in line with Harding and Kloster\cite{harding2024coding}, who estimate the code churn rate (the percentage of code lines that are modified within 2 weeks) to double in 2024 compared to a 2021 baseline from before the wide usage of AI-tools. One developer in the \control{} group mentions: \xquote{Some of my fellow students used the chatbot to write the frontend [...]. Now, [it] is practically unmaintainable because it is just a pile of spaghetti code}{P2}.
Within the same group, another developer had a similar experience, when their team members did not update LLM-generated code to fit their current project's boundaries, resulting in \xquote{[...] others [having] to spend the time saved by [the LLM] to get the code into an acceptable state.}{P4}
The observation was summarized by another developer using the control LLM, pointing out how the \xquote{chatbot was very helpful for general basic understanding, but it reached its limits when it came to more concrete, project-specific things [...]}{P3}. This perception is supported by current work, which often focuses on including the codebase as context in LLM conversations\cite{newton,contextide,llmplugin}, highlighting the gap between the LLM's knowledge of the codebase and the knowledge needed to create code that can be integrated nicely. We argue that, partially, differences of our setting with other studies come from our project-specific setting that does not focus on a single, well-encapsulated tasks, but rather a more realistic open-ended project scenario, which is essential to understand the full range of of an LLM's impact on coding activities.

While we acknowledge that interpretations based on limited data and marginal significance should be made cautiously, the observed trends offer valuable insights for future research. Furthermore, the alignment of our findings regarding modification with participant perceptions and prior work supports the viability of these findings.

\begin{mybox}{}
    To answer \ques{3}, we find that treated LLMs cause more LLM-generated code to be pushed into repositories, with the highest amount in repositories of developers using the \context{} LLM. 
    Despite this disparity, similar amounts of code were changed over time in all groups.
    This highlights how even the more conservative interventions in conversations can increase the LLM's capabilities to generate code that is committed without causing negative trade-offs in that regard, further highlighting the potential of guidance mechanisms.
\end{mybox}

\section{Threats to Validity}

\paragraph{Internal Validity.} The drop-out rate of participants could be a threat to the study's internal validity. This might suggest that participants who continued using the LLMs were more motivated or had a more positive attitude toward them, which particularly threatens conclusions for \ques{2}.
Hence, we collected participants general attitude toward LLM usage at the beginning of the study and found no systematic pattern between participants completing the study and those who drop out. Moreover, since the main development activities were in the first time period, the decreasing numbers may indicate that fewer participants required the help of the LLM.

\paragraph{Statistical Conclusion Validity.} Most participants sent fewer than 200 prompts, except three, one of whom sent over 400. Including them can majorly impact statistical conclusions.
We mitigated this by removing \textit{superusers} during preprocessing to ensure that they are not skewing aggregated statistics.

\paragraph{External Validity.} 
The study population consists of novice developers. 
As the participants were students rather than professional developers, this might limit the generalizability to the full developer population.
However, the study was embedded in a longitudinal, realistic software project resembling conditions developers encounter in their first professional projects.
As such, the study supports viewing the population as novice developers, as is done commonly in Software Engineering research\cite{falessi_empirical_2018}.
Consequently, the results are interpreted with respect to novice developers rather than the educational setting itself.

As novice developers, most of our participants developed their first proper software project during our study, while more experienced developers might have done a lot more projects.
However, a significant portion of developers in practice are junior developers and developers in new positions. 
Thus, while our results may not be generalizable to the entire developer population, they are still relevant to an important subpopulation. 
The goal of this study is to investigate the benefits and harms of interventions in LLM conversations in order to pave the way for new interaction mechanisms.

\paragraph{Ecological Validity.}
The university context threatens the applicability of our results for practical settings and may explain deviations of results from other studies.  
However, conducting a study as controlled as ours, which includes randomly assigning  developers to teams and letting all teams develop the same project for four consecutive months, is infeasible in a realistic setting.
This dedicated focus inherently comes with cost for ecological validity. To increase ecological validity, we designed a realistic project setting, with the project's longevity aligning more closely with real-world development processes, especially compared to studies that focus on small or isolated tasks.

\section{Conclusion}
LLMs have become a well-used tool for developers and a popular research topic.
In this study, we asked whether we can support and guide developers when interacting with LLMs and, more specifically, how such guidance impacts software development.

To this end, we implemented two guidance approaches with minimal interference in the conversation. 
The \context{} treatment supplied the LLM with additional context based on the user's intention, while the \proactive{} treatment offered active guidance through advice, improved prompts, and follow-up questions.
We tested these approaches in a large, longitudinal 4-month study with CS students building a full-stack Web application in their third semester. 

We found that conservative guidance mechanisms in LLMs did not majorly influence participants in their general usage behavior with the LLMs. 
Across the measured outcomes (interaction patterns, self-reported frustration/helpfulness/productivity, and downstream code adoption/modification), we did not observe evidence of negative effects attributable to the interventions. However, we find significant variation in conversation intentions between treatment groups, though the reason for this remains unclear.
In terms of helpfulness, frustration, and productivity, we find that developers receiving the \proactive{} guidance perceived the LLM as the most positive.
The \context{} treatment was perceived critically, but still led to higher usage of LLM-generated code. 

\paragraph{What does it mean?}
We conclude that even such minimal interventions as ours already yield positive effects on developers' satisfaction and how they adopt and modify code. The interventions do not disrupt the workflow compared to the freedom of the control LLM. 
During the course of the project, participants grew more efficient, as conversations grew shorter over time.

Based on this, it is promising to study how higher levels of guidance, such as regular updates of intentions or requesting more context during a conversation, may affect results and user satisfaction. Our initial setup with only sparse intervention of user-LLM interaction can serve as a starting point to better guide (early-career) software developers to efficiently use LLMs in their daily workflow, and, importantly, dispel ethical concerns regarding harmful effects of studies that investigate interventions in user-LLM interactions.

\section*{Statements and Declarations}

\subsection*{Funding}
This work was supported by the Saxon State Ministry for Science, Culture and Tourism (SMWK) through the “Center for Scalable Data Analytics and Artificial Intelligence Dresden/Leipzig” (ScaDS.AI), by the European Social Fund (ESF) together with the German state of Saxony under grant number A.100760692 (Project: Teaching-AI), and by the German Research Foundation DFG (SI 2171/3-2).

\subsection*{Ethical approval}
This study involved voluntary participation of computer science students and did not require formal ethical approval according to the institutional regulations of the participating universities. 
All participants provided informed consent prior to participation. 
Responses were pseudonymized, and no personally identifiable information was published. 
Participation was voluntary, and participants could withdraw at any time without consequence. 
Data was handled in accordance with applicable data protection regulations.

\subsection*{Informed consent}
Participants were informed about the purpose of the study, the voluntary nature of their participation, what data would be collected, how it would be used and protected, and that they could withdraw at any time. 
Data was pseudonymized during analysis and anonymized prior to publication. Informed consent was obtained before any data was collected. Informed consent was obtained prior to data collection.

\subsection*{Author Contributions}
All authors contributed to the study conception and design. 
Material preparation and data collection were performed by Annemarie Wittig and Alina Mailach; Analysis was performed by Annemarie Wittig. The first draft of the manuscript was written by Annemarie Wittig and subsequently revised by Alina Mailach; All authors commented, updated and rephrased previous versions of the manuscript. All authors read and approved the final manuscript.

\subsection*{Data Availability Statement}
The data and material from this study is available in our replication package: 
\url{https://anonymous.4open.science/r/On-the-Prospects-of-Dynamic-LLM-Conversations-in-Software-Development-D22C/README.md}

\subsection*{Conflict of Interest}
The authors have no competing interests to declare that are relevant to the content of this article.

\subsection*{Clinical Trial Number}
Not applicable

\bibliographystyle{spmpsci}
\bibliography{sources}

\end{document}